\documentclass[11pt]{article}
\usepackage{indentfirst,psfig}
\usepackage{apalike}
\newcommand{\bey}[1]{\begin{eqnarray} \label{#1}}
\newcommand{\eey}{\end{eqnarray}}
\newcommand{\beq}[1]{\begin{equation} \label{#1}}
\newcommand{\eeq}{\end{equation}}
\renewcommand{\baselinestretch}{1.0} \tiny\normalsize

\title{
A Statistical Mechanical Approach to Combinatorial Chemistry
}

\author{Michael W. Deem\\
~\\
Chemical Engineering Department\\
University of California\\
Los Angeles, CA 90095--1592}

\begin{document}
\renewcommand{\baselinestretch}{1.0} \tiny\normalsize
\maketitle

\vspace{2in}

To appear in \emph{Advances in Chemical Engineering}.

\newpage
\tableofcontents

\newpage

\section{Introduction}

The goal of combinatorial chemistry is to find compositions
of matter that maximize a specific material property.
When combinatorial chemistry is applied to materials discovery,
the desired property may be superconductivity, magnetoresistance,
luminescence, ligand specificity, sensor response, or catalytic
activity.  
When combinatorial chemistry is applied to proteins, the
desired property may be enzymatic activity, fluorescence,
antibiotic resistance, or substrate binding specificity.
In either case, the property to be optimized, the
figure of merit, is generally an unknown function of the
variables and can be measured only experimentally.

Combinatorial chemistry is, then, a search over a multi-dimensional
space of composition and
non-composition variables for regions of high figure of merit.
A traditional synthetic chemist would carry out this search
by using chemical intuition  to synthesize a few
initial molecules.  
Of these molecules, those that have a favorable figure of merit
would be identified.
A homologous series of compounds similar to those
best starting points would then be synthesized.
Finally, of the compounds in these homologous series, that with the best
figure of merit
would be identified as the optimal material.

If the space of composition and non-composition variables
is sufficiently large, novel, or unfamiliar, the traditional synthetic
approach may lead to the identification of materials that are not truly
the best.  It is in this case that combinatorial chemistry becomes useful.
In combinatorial chemistry, trial libraries of molecules are synthesized
instead of trial molecules.   By synthesizing and screening for figure of
merit an entire library of $10^2 - 10^5$ molecules instead of a single
molecule, the variable space can be searched much more thoroughly.
In this sense, combinatorial chemistry is a natural extension of
traditional chemical synthesis. Intuitive determination of the individual
molecules to synthesize is replaced by methods
for design of the molecular libraries.
Likewise, synthesis of homologous compounds is replaced by redesign of the
libraries for multiple rounds of parallel screening experiments.

While the combinatorial approach
attempts to search composition space broadly,
an exhaustive search is usually not possible.  It would take,
for example, a library of $9 \times 10^6$ compounds to search
a five-component system at a mole fraction resolution of 1\%.
Similarly, it would take a library of $20^{100} \approx 10^{130}$ proteins
to search exhaustively the space of all 100 amino-acid protein domains.
Clearly, a significant aspect to the design of a combinatorial chemistry
experiment is the design of the library.  The library members should
be chosen so as to search the space of variables as effectively as possible,
given the experimental constraints on the library size.

The task of searching composition space in combinatorial chemistry
for regions of high figure of merit
is very similar to the task of searching configuration space by Monte
Carlo computer simulation for regions of
low free energy.  The space searched by Monte Carlo computer simulation
is often extremely large, with $10^4$ or more continuous dimensions.
Yet, with recent advances in the design of Monte Carlo algorithms, 
one is able to locate reliably the regions of low free energy even for fairly
complicated molecular systems.

This chapter pursues the analogy between combinatorial chemistry and
Monte Carlo computer simulation.  Examples of how to design libraries
for both 
materials discovery 
and 
protein molecular evolution 
will be given.  
For materials discovery, 
the concept of library redesign, or the use previous experiments
to guide the design of new experiments, will be introduced.
For molecular evolution, 
examples of how to use ``biased'' Monte Carlo to search
the protein sequence space will be given.  
Chemical information, whether intuition, theoretical calculations,
or database statistics, can be naturally incorporated 
as an \emph{a priori} bias in the Monte Carlo approach to
library design in combinatorial chemistry.
In this sense, combinatorial chemistry can be viewed as an
extension of traditional chemical synthesis.

\section{Materials Discovery}

A variety of materials have been optimized or developed to date by
combinatorial methods.  Perhaps the first experiment to gather
great attention was the demonstration that
inorganic oxide high-$T_{\rm c}$ superconductors could be identified
by combinatorial methods \cite{combi_4}.  By searching several 128-member libraries of
different inorganic oxide systems, the known compositions of
superconducting
BiSrCaCuO and YBaCuO were identified.  Since then, many demonstrations of
finding known materials and discoveries of new materials have
appeared.  Known compositions of
giant magnetoresistant materials have been identified in
libraries of various cobalt oxides \cite{combi_22}.
Blue and red phosphors have been identified
from large libraries of 25000 different inorganic oxides
\cite{combi_11,combi_5,combi_10}.
Polymer-based sensors for various organic vapors have been
identified by combinatorial methods \cite{combi_24}.
Catalysts for the oxidation of CO to CO$_2$ have been
identified by searching ternary compounds of Pd, Pt, and Rh or
Rh, Pd, and Cu \cite {combi_15,combi_7}.  Phase diagrams of zeolitic materials
have been mapped out by a combinatorial ``multiautoclave'' \cite{combi_17}.
Novel enantioselective catalysts have been found by searching
libraries of transition metal-peptide complexes \cite{combi_18}.
Novel phosphatase catalysts were found by searching libraries of
carboxylic acid-functionalized polyallylamine  polymers \cite{combi_16}.
New catalysts and conditions for C-H insertion have been
found by screening of ligand-transition metal systems \cite{combi_31}.
A new catalyst for the conversion of methanol in a direct methanol
fuel cell was identified by searching the quaternary composition space
of Pt, Ir, Os, and Ru \cite{combi_21}.  Finally,
a novel thin-film high-dielectric compound that may be used in future
generation of DRAM chips
was identified by searching through
over 30 multicomponent, ternary oxide systems \cite{combi_8}.

The task of identifying the optimal compound in 
a materials discovery  experiment
can be reformulated
as one of searching a multi-dimensional space, with the material
composition, impurity levels, and synthesis conditions as variables.
Present approaches to combinatorial library design and screening
invariably perform a grid search in composition space, followed by a
``steepest-ascent'' maximization of the figure of merit.  This
procedure becomes inefficient in high-dimensional spaces or when
the figure of merit is not a smooth function of the variables.
\emph{Indeed, the
use of a grid search is what has limited essentially all current
combinatorial chemistry experiments to quaternary compounds, \emph{i.e.}\ to
searching a space with three variables.}
What is
needed is an automated, yet more efficient, procedure
for searching composition space.

An analogy with the computer simulation technique of Monte Carlo
allows us to design just such an efficient protocol
for searching the variable space 
\cite{Deem2000}.
In materials discovery, a search is made through the composition and
non-composition
 variables to find good figure-of-merit values.  In Monte Carlo,
a search is made through configuration space to find
regions of low free energy.
By using insight gained from the design of Monte Carlo methods,
the search in materials discovery can be improved.

\subsection{The Space of Variables}
Several variables can be manipulated in order to seek the
material with the optimal figure of merit.  Material composition is
certainly a variable.  But also, film thickness \cite{combi_8} and deposition
method \cite{Novet95}
are variables for materials made in thin film form. The
processing history, such as temperature, pressure, pH, and atmospheric
composition, is a variable.  The guest composition or impurity level
can greatly affect the figure of merit \cite{combi_7}. In addition, the
``crystallinity'' of the material can affect the observed figure of
merit \cite{combi_8}. Finally, the method of nucleation or synthesis may
affect the phase or morphology of the material and so affect the
figure of merit \cite{Davis95,Zones98}.

There are important points to note about these variables.
First, a small impurity composition can cause a big change in the
figure of merit, as
seen by the rapid variation of catalytic activity in
the  Cu/Rh oxidation catalyst \cite{combi_7}.
Second, the phases in thin film are not necessarily the same as those
in bulk, as seen in the case of the
thin-film  dielectric, where the optimal material was
found outside the region where the bulk phase forms \cite{combi_8}.
Finally, the ``crystallinity'' of the material can affect the
observed figure of merit, again as seen in the thin-film
dielectric example \cite{combi_8}.

\subsection{Library Design and Redesign}

The experimental challenges in combinatorial chemistry
appear to lie mainly in the screening methods and in the
technology for the creation of the libraries.
The theoretical challenges, on the other hand,
appear to lie mainly in the library design and redesign strategies.  
It is this second question that
is addressed by the 
analogy with Monte Carlo computer simulation.

Combinatorial chemistry differs from usual Monte Carlo simulations
in that several simultaneous searches of the variable space are
carried out.
That is, in a typical combinatorial chemistry experiment, several samples,
\emph{e.g.}\ 10000, are synthesized and screened for figure of merit
at one time.  With the results of this first round, a new set of
samples can be synthesized and screened.  This procedure can be
repeated for several rounds, although current materials
discovery experiments have
not systematically exploited this feature.

  Pursuing the analogy
with Monte Carlo, each round of combinatorial chemistry corresponds
to a move in a Monte Carlo simulation.  Instead of tracking one
system with many configurational degrees of freedom, however,
many samples are tracked, each with several composition 
and non-composition degrees of freedom.  
Modern experimental technology is what allows for
the cost-effective synthesis and screening of
multiple sample compositions.

The technology for materials discovery is still in the
developmental stage, and future progress can still be
influenced by theoretical considerations.
In this spirit, I assume that  the composition and
non-composition variables  of each
sample can be changed independently, as in
spatially addressable libraries \cite{combi_17,combi_2a}.
This is significant, because it allows
great flexibility in how the space can be searched with
a limited number of experimental samples.

\begin{figure}[tbp]
\begin{minipage}[h]{4.00in}
\psfig{file=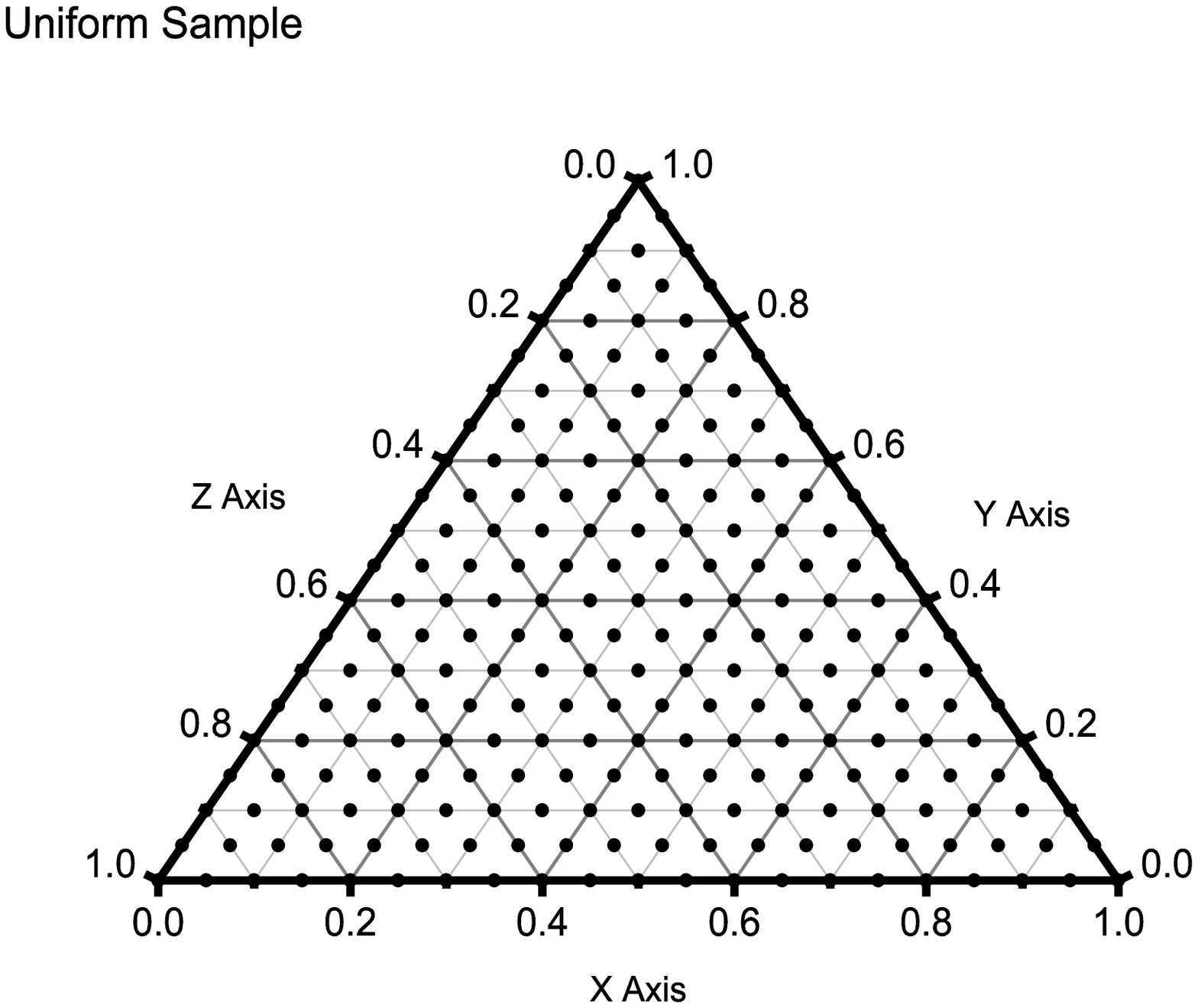,width=2.0in}
\end{minipage}
\hfill
\begin{minipage}[h]{4.00in}
\psfig{file=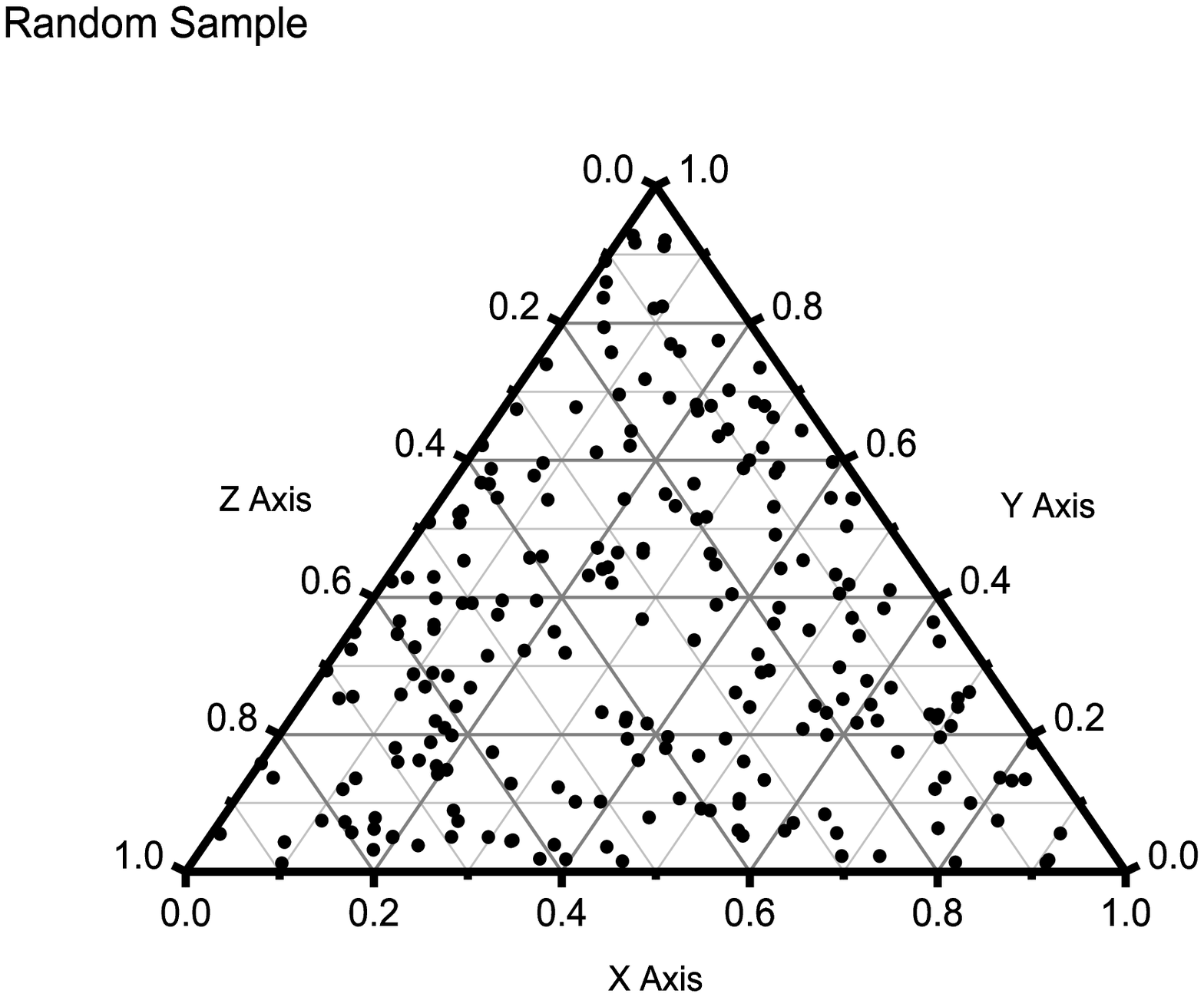,width=2.0in}
\end{minipage}
\hfill
\begin{minipage}[h]{4.00in}
\psfig{file=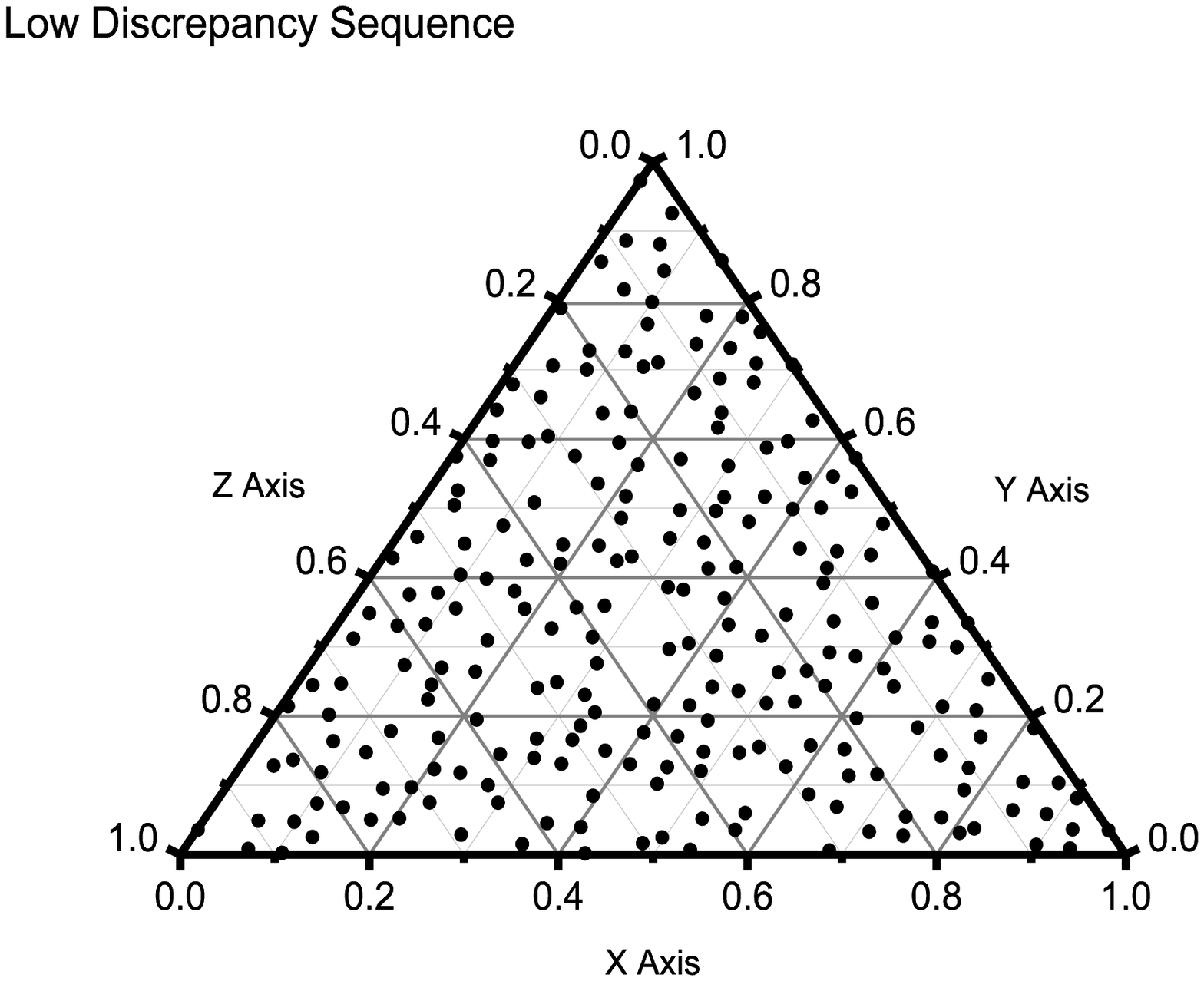,width=2.0in}
\end{minipage}
\caption{
Shown are the grid, random, and low-discrepancy sequence
approaches to designing the first library in a
materials discovery experiment with three compositional variables.
The random approach breaks the regular pattern of the grid search, and
the low-discrepancy sequence approach avoids overlapping points
that may arise in the random approach.
}
\label{fig:random}
\end{figure}

Current experiments uniformly tend to
perform a grid search on the
composition and non-composition variables.
It is preferable, however, to choose the
variables statistically
 from the allowed values.  It is also possible  to consider
choosing the variables in a fashion that attempts to maximize the amount
of information gained from the limited number of samples screened,
{\em via} a quasi-random, low-discrepancy sequence \cite{LDS2,LDS1}.
Such sequences attempt to eliminate the redundancy that naturally occurs
when a space is searched statistically, and they have several
favorable theoretical properties.  An illustration of these
three approaches to materials discovery library design
is shown in Figure \ref{fig:random}.

Information about the figure-of-merit landscape in the composition
and non-composition variables can be incorporated
by multiple rounds of screening.
One convenient way to incorporate this
feedback as the experiment proceeds is by
treating the
combinatorial chemistry experiment as a Monte Carlo in the laboratory.
This approach leads to sampling the experimental figure of merit, $E$,
proportional to $\exp(\beta E)$.
If $\beta$ is large, then the Monte Carlo procedure
will seek out values of the composition and non-composition variables
that maximize the figure of merit.  If $\beta$ is too large, however,
the Monte Carlo procedure will get stuck in relatively low-lying local
maxima.  The first round is initiated by choosing the
composition and non-composition
variables statistically from the allowed values.
The variables are changed in succeeding rounds as dictated by the Monte Carlo
procedure.

Several general features of the method for changing the 
variables can be enumerated.  The statistical
method of changing the variables can be biased by concerns such
as material cost, theoretical or experimental \emph{a priori} 
insight into how the figure of merit is likely to change, and
patentability.   Both the composition and non-composition variables
will be changed in each round.  Likely, it would be desirable
to have a range of move sizes for both types of variables.  
The characteristic move size would likely best be determined by
fixing the acceptance ratio of the moves, as is customary in 
Monte Carlo simulations \cite{Frenkel_book}.  In addition,
there would likely be a smallest variable change that would be
significant, due to experimental resolution limitations in the screening step.
Finally, a steepest-ascent
optimization to find the best local optima of the figure of
merit would likely be beneficial
at the end of a materials discovery experiment driven by such a 
Monte Carlo strategy.

\subsection{Searching the Variable Space by Monte Carlo}
Two ways of changing the variables are considered:
a small random change of the variables of a randomly
chosen sample
and a swap of a subset of
the variables between two randomly chosen samples.  
Swapping is useful when there is a hierarchical structure
to the variables.  The swapping event allows for the combination of
beneficial subsets of variables between different samples.
For example, a good set of composition variables might be
combined with a particularly good impurity composition.
Or, a good set of composition variables might be
combined with a good set of processing variables.
These moves are
repeated until all the samples in a round have been modified.
The values of the figure of merit for the
proposed new samples are then measured.
Whether to accept the newly proposed
samples or to keep the current samples for the next round is decided
according to the
detailed balance acceptance criterion.  For a
random change of one sample, the Metropolis acceptance probability
is applied:
\begin{equation}
p_{\rm acc}(c \to p) = \min\left\{1,
\exp\left[\beta \left(E_{\rm proposed} - E_{\rm current }\right)\right]\right\} \ .
\label{2}
\end{equation}
  Proposed samples that
increase the figure of merit are always accepted; proposed samples that
decrease the figure of merit are accepted with the Metropolis
probability.  Allowing the figure of merit occasionally to decrease is
what allows samples to escape from local maxima.  
Moves that lead to invalid values of the composition or
non-composition variables are rejected.

\begin{figure}[tbp]
\centering
\leavevmode
\psfig{file=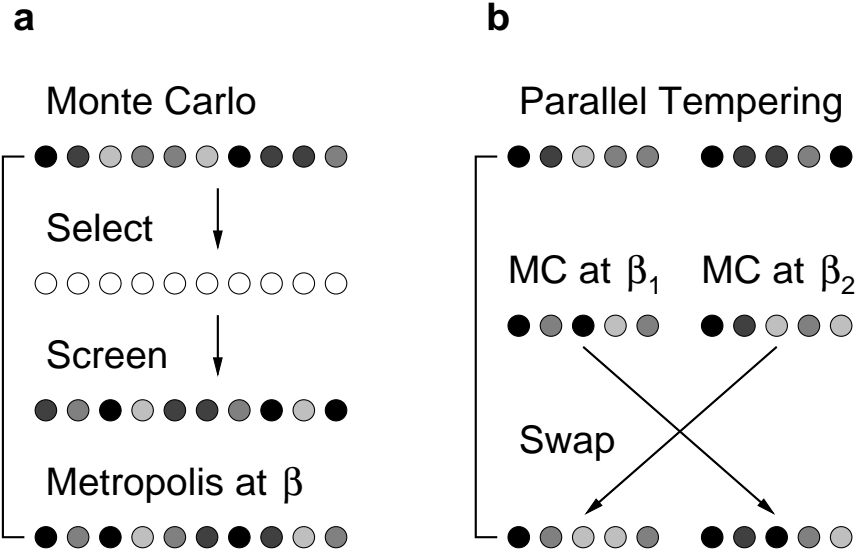,clip=,width=4.5in}
\caption{
Schematic of the Monte Carlo library design and redesign
strategy, from \cite{Deem2000}.
a) One Monte Carlo round with 10 samples.  Shown are an
initial set of samples, modification of the samples, 
measurement of the new figures of merit, and
the Metropolis criterion for acceptance or rejection of the new samples.
b) One parallel tempering round with 5 samples at $\beta_1$ and
5 samples at $\beta_2$.  In parallel tempering, several
Monte Carlo simulations are performed at different temperatures,
with the additional possibility of sample
exchange between the simulations at different temperatures.
}
\label{fig:mc}
\end{figure}

For the swapping move applied to samples $i$ and $j$,
the modified acceptance probability is applied:
\begin{eqnarray}
p_{\rm acc}(c \to p) &=& \min\bigg\{1,
\exp\bigg[\beta \bigg(E_{\rm proposed}^i + E_{\rm proposed}^j
\nonumber \\
&&~~~~~~~~~~~~~~ - E_{\rm current}^i - E_{\rm current}^j \bigg)\bigg]\bigg\} \ .
\label{3}
\end{eqnarray}
Figure \ref{fig:mc}a
shows one round of a Monte Carlo procedure.  The parameter $\beta$
is not related to the thermodynamic temperature of the
experiment and
should be optimized for best efficiency.  The characteristic sizes of
the random changes in the composition and non-composition variables are also
parameters that should be optimized.

If the number of composition and non-composition variables is too great, or if
the figure of merit changes with the variables in a too-rough fashion,
normal Monte Carlo will not achieve effective sampling.  Parallel
tempering is a natural extension of Monte Carlo that is used to study
statistical \cite{Geyer}, spin glass \cite{Parisi}, and molecular \cite{Deem1}
systems with rugged energy landscapes.  Our most powerful
protocol incorporates the method of parallel tempering for changing
the system variables.  In parallel tempering, a fraction of
the samples are updated
 by Monte Carlo with parameter $\beta_1$, a fraction by
Monte Carlo with parameter $\beta_2$, and so on.  At the end of each
round, samples are randomly exchanged between the groups with
different $\beta$'s, as shown in Figure \ref{fig:mc}b.  
The acceptance probability for exchanging two samples is
\begin{equation}
p_{\rm acc}(c \to p) = \min\left\{1,
\exp\left[-\Delta \beta \Delta E \right]\right\} \ ,
\label{4}
\end{equation}
where $\Delta \beta$ is the difference in the values of
 $\beta$ between the two groups,
and $\Delta E$ is the difference in the figures of merit between the
two samples.
It is important to
notice that this exchange step does not involve any extra screening compared
to Monte Carlo and is, therefore, ``free'' in terms of
experimental costs.  This step is, however, dramatically effective at
facilitating the protocol to escape from local maxima.  The number of
different systems and the temperatures of each system are parameters
that must be optimized.

To summarize, the first round of combinatorial chemistry consists of
the following steps:
constructing the  initial library of samples,
measuring the initial figures of merit,
changing the variables of each sample a small random amount or
swapping subsets of the variables between pairs of samples,
constructing the proposed new library of samples,
measuring the figures of merit of the proposed new samples, 
accepting or rejecting each of the proposed new samples, and
performing parallel tempering exchanges.
Subsequent rounds of combinatorial chemistry repeat these steps,
starting with making changes to the current values
of the composition and non-composition
 variables.  These steps are repeated for as many rounds
as desired, or until maximal figures of merit are found.

\subsection{The Simplex of Allowed Compositions}

The points to be sampled in materials discovery are the allowed
values of the composition and non-composition variables.  Typically, the
composition variables are specified by the mole fractions.  Since the
mole fractions sum to one, sampling on these variables requires special
care.

\begin{figure}[tbp]
\begin{minipage}[h]{4.00in}
\psfig{file=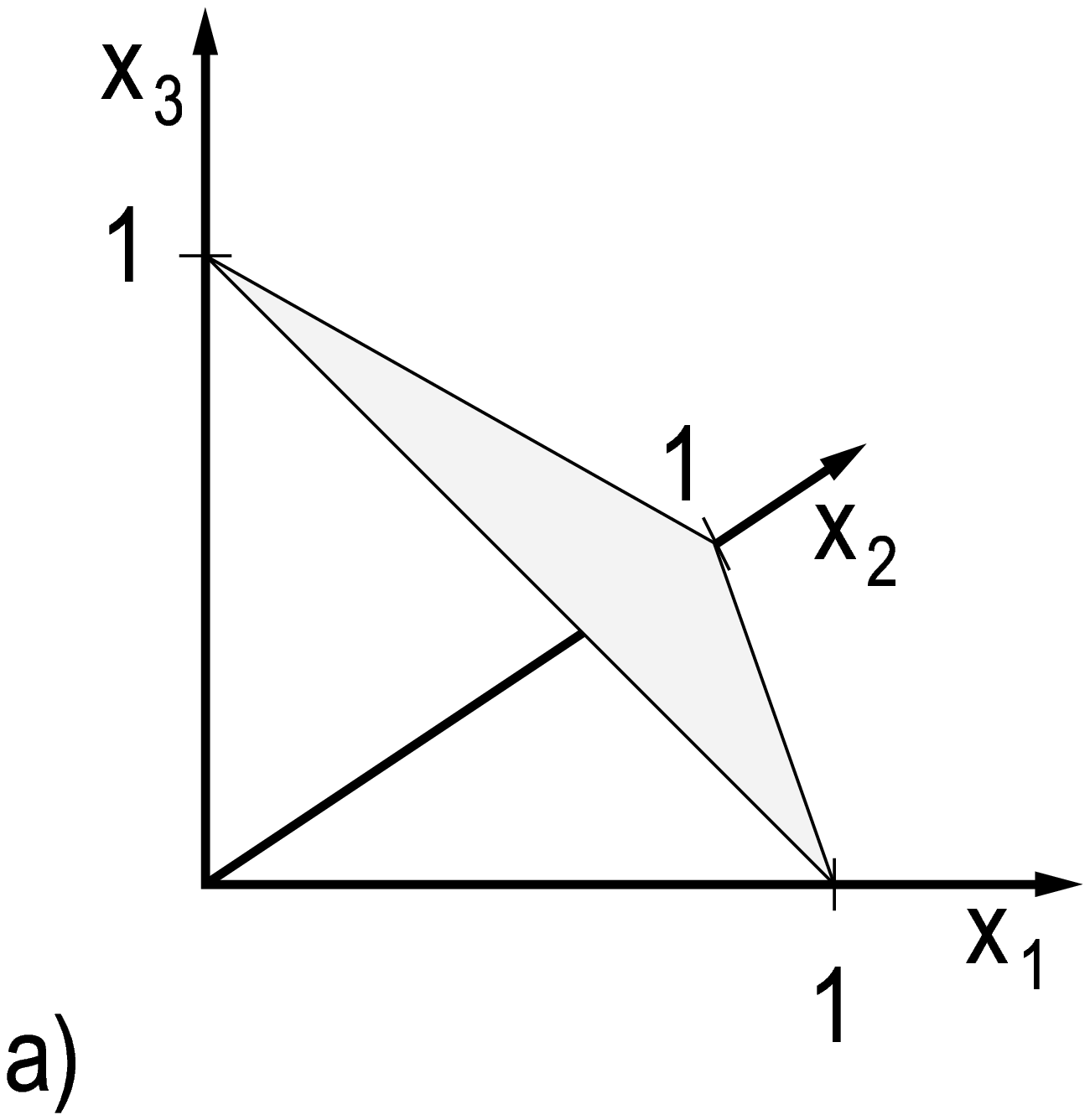,width=2.0in}
\end{minipage}
\hfill
\begin{minipage}[h]{4.00in}
\psfig{file=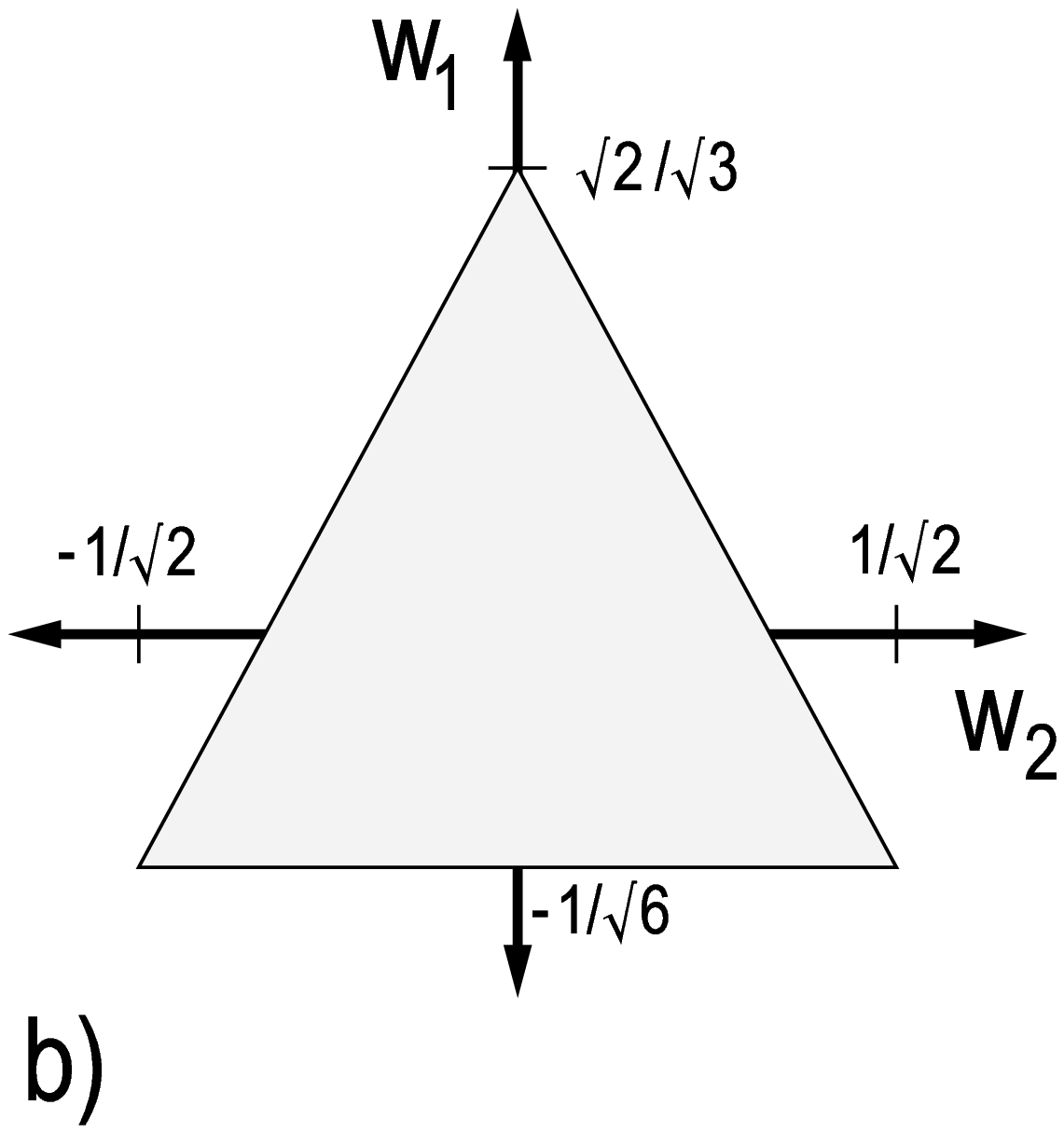,width=2.0in}
\end{minipage}
\caption{
The allowed composition range of a three-component system 
is shown in the a) original composition
variables, $x_i$, and b) Gram-Schmidt variables, $w_i$.
}
\label{fig:simplex}
\end{figure}

In particular, the specification or modification of the
the $d$ mole fraction
variables, $x_i$, is done in the $(d-1)$-dimensional hyper-plane
orthogonal
to the $d$-dimensional vector
$(1,1,\ldots,1)$.  This procedure ensures that the constraint  $\sum_{i=1}^d x_i = 1$
is maintained.  This subspace is identified by a
Gram-Schmidt procedure, which 
identifies a new set of basis vectors, $\{{\bf u}_i \}$, that
span this hyper-plane.  Figure \ref{fig:simplex} illustrates
the geometry for the case of three composition variables.

The new basis set is identified as follows.  First, ${\bf u}_d$ is defined
to be the unit vector orthogonal to the allowed hyper-plane:
\beq{200}
{\bf u}_d = \left(\frac{1}{\sqrt d },
       \frac{1}{\sqrt  d},
       \ldots, 
       \frac{1}{\sqrt d } \right) \ .
\eeq
The remaining ${\bf u}_i, 1 \le i < d,$
are chosen to be orthogonal to ${\bf u}_d$,
 so that
they lie in the allowed hyper-plane.  Indeed, the ${\bf u}_i$ form an
orthonormal basis for the composition space.  This orthonormal basis
is identified by the Gram-Schmidt procedure.  First, the original
composition  basis vectors are defined:
\bey{201}
{\bf e}_1 &=& \left(1, 0, \ldots, 0, 0 \right)
\nonumber \\
{\bf e}_2 &=& \left(0, 1, \ldots, 0, 0  \right)
\nonumber \\
&\vdots&
\nonumber \\
{\bf e}_{d-1} &=& \left(0, 0, \ldots, 1, 0 \right) \ .
\eey
Each ${\bf u}_i$ is identified by projecting these basis
vectors onto the space orthogonal to 
${\bf u}_d$ and
the ${\bf u}_j, j<i,$ already
identified:
\bey{202}
{\bf u}_1 &=& 
\frac
{
{\bf e}_1 - ({\bf e}_1 \cdot {\bf u}_d) {\bf u}_d
}
{
\vert {\bf e}_1 - ({\bf e}_1 \cdot {\bf u}_d) {\bf u}_d \vert
}
\nonumber \\
{\bf u}_2 &=& 
\frac
{
{\bf e}_2 - ({\bf e}_2 \cdot {\bf u}_d) {\bf u}_d
          - ({\bf e}_2 \cdot {\bf u}_1) {\bf u}_1
}
{
\vert
{\bf e}_2 - ({\bf e}_2 \cdot {\bf u}_d) {\bf u}_d
          - ({\bf e}_2 \cdot {\bf u}_1) {\bf u}_1
\vert
}
\nonumber \\
&\vdots&
\nonumber \\
{\bf u}_i &=& 
\frac
{
{\bf e}_i - ({\bf e}_i \cdot {\bf u}_d) {\bf u}_d
          - \sum_{j=1}^{i-1} ({\bf e}_i \cdot {\bf u}_j) {\bf u}_j
}
{
\vert
{\bf e}_i - ({\bf e}_i \cdot {\bf u}_d) {\bf u}_d
          - \sum_{j=1}^{i-1} ({\bf e}_i \cdot {\bf u}_j) {\bf u}_j
\vert
}
\eey
A point in the allowed composition range is specified by the
vector ${\bf x} = \sum_{i=1}^d w_i {\bf u}_i$, with
$w_d = 1/\sqrt d$.
Note that the values $w_i$ are related to the composition
values $x_i$ by a rotation matrix,
since the Gram-Schmidt procedure simply identifies a
rotated basis for the composition space:
\beq{203}
{\bf x} = R {\bf w} \ ,
\eeq
 where
$R_{ij}$ is given by the $i$-th component of ${\bf u}_j$.
Each of the numbers $w_i, 1 \le i < d,$ 
is to be varied in the materials discovery experiment.
Not all values of $w_i$
are feasible, however, since the constraint
$x_i \ge 0$ must be satisfied.  Feasible values are identified by
transforming the $w_i$ to the $x_i$ by Eq.\ (\ref{203}), and then checking that
the composition variables are non-negative.  The constraint
that the composition variables sum to unity is automatically
ensured by the choice $w_d = 1/\sqrt d$.

\subsection{Significance of Sampling}
Sampling the figure of merit by Monte Carlo, rather than
global optimization by some other method, is favorable
for several reasons.
First, Monte Carlo is
an effective stochastic optimization method.
Second, simple global optimization may be misleading since
concerns such as patentability, cost of materials, and ease of synthesis
are not usually included in the experimental figure of merit.
Moreover, the
screen that is most easily performed in the laboratory, the ``primary
screen,'' is usually only roughly correlated with the true figure of
merit.  Indeed, after finding materials that look
promising based upon the primary screen, experimental secondary and
tertiary screens are usually performed to identify that material which is
truly optimal.  Third, it might be advantageous to screen for several
figures of merit at once. For example, it might be profitable to
search for reactants and conditions that lead to the synthesis of
several zeolites with a particularly favorable property, such as the
presence of a large pore.
As another example, it might be useful to search for several
electrocatalysts that all possess a useful property, such as
being able to serve as the anode or cathode material
in a particular fuel cell.

For all of these reasons, sampling
by Monte Carlo to produce several candidate materials is
preferred over global optimization.

\subsection{The Random Phase Volume Model}

The ultimate test of new, theoretically-motivated protocols for
materials discovery is, of course, experimental.  In order to
motivate such experimentation, 
the effectiveness of these protocols is demonstrated by 
combinatorial chemistry experiments where the experimental
screening step is replaced by figures of merit returned by
the Random Phase Volume Model.
The Random Phase Volume Model is not fundamental to the protocols;
it is introduced as a simple way to test, parameterize, and
validate the various searching methods.

The Random Phase Volume Model relates 
the figure of merit to the composition and non-composition
variables in a statistical way.  The model is fast enough to allow for
validation of the proposed searching methods on an enormous number of
samples, yet possesses the correct statistics for the figure-of-merit
landscape.

\begin{figure}[tbp]
\centering
\leavevmode
\psfig{file=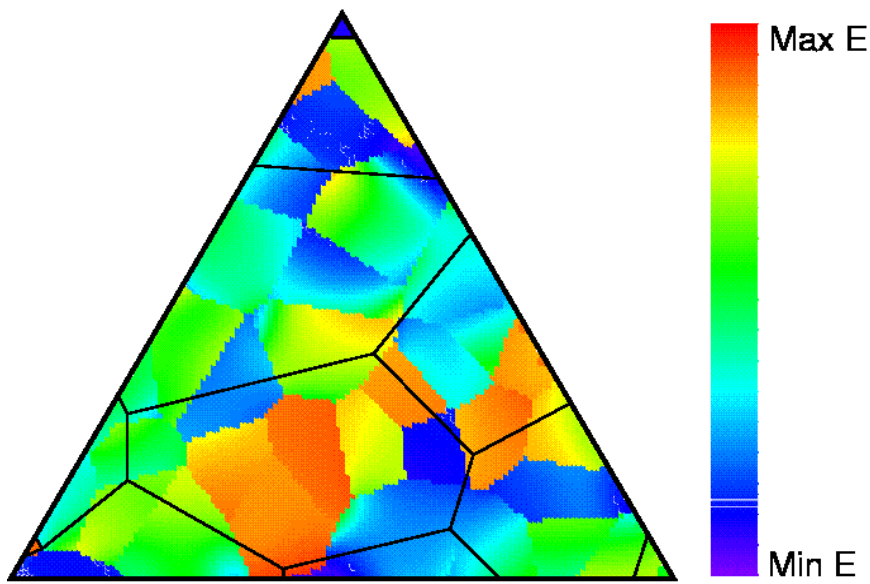,clip=,width=4.5in}
\caption{
The Random Phase Volume Model, from
\cite{Deem2000}.  The model is shown for the
case of three composition variables and one non-composition variable.
The boundaries of the ${\bf x}$ phases are evident by the
sharp discontinuities in the figure of merit.  To generate this
figure, the ${\bf z}$ variable was held constant.  The boundaries of
the ${\bf z}$ phases are shown as thin dark lines.}
\label{fig:phases}
\end{figure}

The composition mole fractions are non-negative and
sum to unity, and so the allowed compositions are constrained to lie within a 
simplex in $d-1$ dimensions.  For the familiar ternary system, this
simplex is an equilateral triangle, as shown in Figure \ref{fig:simplex}b.
Typically, several phases will exist for different compositions
of the material.  The figures of merit will be dramatically different 
between each of these distinct phases.  To mimic this expected behavior,
the composition variables are grouped in the Random Phase Volume Model 
into phases centered around $N_x$ points ${\bf x}_\alpha$
randomly placed within the allowed composition range. The phases form
a Voronoi diagram \cite{Sedgewick}, as shown in Figure \ref{fig:phases}.  

The Random Phase Volume Model is defined for any number of composition
variables, and the number of phase points is defined by requiring the
average spacing between phase points to be $\xi = 0.25$.  To avoid
edge effects, additional points are added in a belt of width $2 \xi$
around the simplex of allowed compositions.
The number of phase points for different grid spacings is
shown in Table \ref{table:grid}.
\begin{table}[tbp]
\caption{Number of phase points as a function of dimension and
spacing.}
\label{table:grid}
\begin{center}
\begin{tabular}[t]{r@{.}l lr c  r@{.}l lr}
\hline
\hline
\multicolumn{2}{c}{$\xi$} & $d$ & Number of points &
\hspace{1in}&
\multicolumn{2}{c}{$\xi$} & $d$ & Number of points\\
\hline
0&1 & 3   & 193   && 0&3 & 3   & 59       \\
0&1 & 4   & 1607  && 0&3 & 4   & 353      \\
0&1 & 5   & 12178 && 0&3 & 5   & 2163     \\
0&1 & 6   & 81636 && 0&3 & 6   & 12068    \\
0&2 & 3   & 86    && 0&35 & 3  & 53       \\
0&2 & 4   & 562   && 0&35 & 4  & 306      \\
0&2 & 5   & 3572  && 0&35 & 5  & 1850     \\
0&2 & 6   & 20984 && 0&35 & 6  & 10234    \\
0&25 & 3  & 70       \\
0&25 & 4  & 430      \\
0&25 & 5  & 2693     \\
0&25 & 6  & 15345    \\
\hline
\hline
\end{tabular}
\end{center}
\end{table}

The figure of merit should
change dramatically between composition phases.
Moreover, within each phase
$\alpha$, the figure of merit should also vary with ${\bf y} = {\bf x}
- {\bf x}_\alpha$ due to crystallinity effects such as crystallite
size, intergrowths, defects, and faulting \cite{combi_8}. In addition, the
non-composition variables should
 also affect the measured figure of merit.  The
non-composition variables are denoted by the $b$-dimensional vector
${\bf z}$, with each component constrained to fall within the range
$[-1,1]$ without loss of generality.  
There can be any number of non-composition variables. 
The figure of merit depends on the composition and non-composition variables
in a correlated fashion.  In particular, how the figure of merit changes
with the non-composition variables should depend on the values of the
composition variables.  To mimic this behavior within the Random Phase Volume
Model,  the non-composition variables 
also  fall within $N_z$ non-composition phases
defined in the space of composition variables.  There are
a factor of 10 fewer non-composition phases than
composition phases.  

The
functional form of the model when ${\bf x}$ is in 
composition phase $\alpha$ 
and non-composition phase $\gamma$ is
\begin{eqnarray}
&&E({\bf x}, {\bf z}) = 
\nonumber \\
&&U_\alpha 
+ \sigma_x
\sum_{k=1}^q  \sum_{i_1 \ge \ldots \ge i_k = 1}^d
f_{i_1 \ldots i_k} \xi_x^{-k} A^{(\alpha k)}_{i_1 \ldots i_k} \, 
y_{i_1} y_{i_2} \ldots  y_{i_k}
\nonumber \\
&& +  \frac{1}{2} \left( W_\gamma  
+
\sigma_z
\sum_{k=1}^q  \sum_{i_1 \ge  \ldots \ge i_k = 1}^b
f_{i_1 \ldots i_k} \xi_z^{-k} B^{(\gamma k)}_{i_1 \ldots i_k} 
z_{i_1} z_{i_2} \ldots  z_{i_k} \right)  \ ,
\nonumber \\
\label{rpvm}
\end{eqnarray}
where $f_{i_1 \ldots i_k}$ is a constant symmetry factor, $\xi_x$ and
$\xi_z$ are constant scale factors, and
$U_\alpha$, $W_\gamma$, $A^{(\alpha k)}_{i_1 \ldots i_k}$, and
$B^{(\gamma k)}_{i_1 \ldots i_k}$ are random Gaussian variables with unit
variance. In more detail, the symmetry factor is given by
\begin{equation}
f_{i_1 \ldots i_k} = \frac{k!}{ \prod_{i=1}^l o_i!} \ ,
\label{6}
\end{equation}
 where $l$
is the number of distinct integer values  in the
set $\{i_1, \ldots, i_k\}$, and $o_i$
is the number of times that distinct value $i$ is repeated in the set.
Note that $ 1 \le l \le k$ and
$\sum_{i=1}^l o_i = k$.
The scale factors are chosen so that each term in the multinomial
contributes roughly the same amount: $\xi_x = \xi/2$ and $\xi_z = 
(\langle z^6 \rangle / \langle z^2 \rangle)^{1/4} =(3/7)^{1/4}$.
The $\sigma_x$ and $\sigma_z$ are chosen so that the
multinomial, crystallinity
 terms contribute 40\% as much as the constant, phase terms on
average. For both multinomials $q=6$.  
 As Figure \ref{fig:phases}
shows, the Random Phase Volume Model describes a rugged
figure-of-merit landscape, with subtle variations, local maxima, and
discontinuous boundaries.

\subsection{Several Monte  Carlo Protocols}
Six different ways of searching the variable space
are tested with increasing
numbers of composition and non-composition variables.
The total number of samples whose figure of merit will be measured is
fixed at $M=
100000$, so that all protocols have the same experimental cost.
The single pass protocols grid, random, and 
low-discrepancy sequence (LDS)
are considered.  
For the grid method, the number of samples in the composition
space is $M_x = M^{(d-1)/(d-1+b)}$ and the number of samples
in the non-composition space is $M_z =
M^{b/(d-1+b)}$.  The grid spacing of the composition variables
is
$\zeta_x = \left( V_d/M_x \right)^{1/(d-1)}$, where
\begin{equation}
V_d = \frac{\sqrt d} {(d-1)!}
\label{7}
\end{equation}
is the volume of the allowed composition simplex. Note that the
distance from the centroid of the simplex to the closest point on the
boundary of the simplex is 
\begin{equation}
R_d = \frac{1}{\left[d (d-1)\right]^{1/2}} \ .
\label{8}
\end{equation}
The spacing for each
component of the non-composition
 variables is $\zeta_z = 2/M_z^{1/b}$.
 For the LDS method, different quasi-random sequences are used for the
composition and non-composition variables.  
The feedback protocols Monte
Carlo, Monte Carlo with swap, and parallel tempering are 
considered.
The Monte Carlo parameters were optimized
on test cases.  It was optimal to perform
100 rounds of 1000 samples with $\beta = 2$ for $d=3$ and
$\beta = 1$ for $d=4$ or 5, and
$\Delta x = 0.1 R_d$ and $\Delta z = 0.12$ for the maximum random
displacement in each component.
The swapping move consisted of an attempt to swap all of the non-composition
values between the two chosen samples, and it was optimal to use
$P_{\rm swap} \simeq 0.1$ for the
probability of a swap versus a regular random displacement.
 For parallel tempering it was optimal to perform 100 rounds with 1000
samples, divided into three subsets:
50 samples at $\beta_1 = 50$, 500 samples at $\beta_2 = 10$, and 450 samples
at $\beta_3 = 1$.  The 50 samples at large $\beta$ essentially
perform a ``steepest-ascent'' optimization
 and have smaller $\Delta x = 0.01 R_d$
and $\Delta z = 0.012$.

\subsection{Effectiveness of the Monte Carlo Strategies}

\begin{figure}[tbp]
\centering
\leavevmode
\psfig{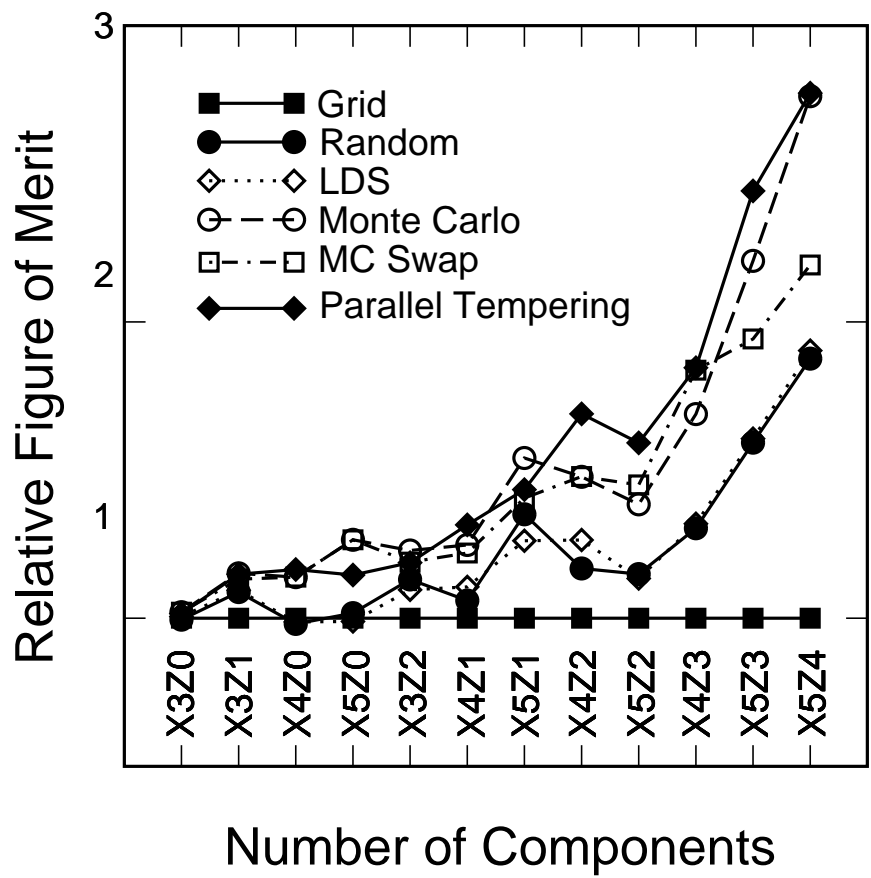}
\caption{The maximum figure of merit found with different protocols on
systems with different number of composition ({\bf x}) and non-composition
({\bf z}) variables,
from \cite{Deem2000}.
The results are scaled to the maximum found by
the grid searching method. 
Each value is averaged over scaled results on 10
different instances of the Random Phase Volume Model with different
random phases.  
The Monte Carlo methods are especially
effective on the systems with larger number of variables, where the
maximal figures of merit are more difficult to locate.
}
\label{fig:results}
\end{figure}

The figures of merit found by the protocols are 
shown in Figure \ref{fig:results}. 
The single-round protocols, random and low-discrepancy sequence, find
better solutions than does grid in one round of experiment.
Interestingly, the low-discrepancy sequence
approach fares no better than does random,
despite the desirable theoretical properties of low-discrepancy
sequences.

The multiple-round, Monte Carlo protocols appear to be especially
effective on the more difficult systems with larger numbers of 
composition and non-composition variables.  
That is, the Monte Carlo methods have a tremendous advantage
over one pass methods, especially as the number of variables
increases, with parallel tempering the best method. 
The Monte Carlo methods, in essence, gather more information about
how best to search the variable space with each succeeding round.
This feedback mechanism proves to be effective even for the
relatively small total sample size of 100000 considered here.
It is expected that the advantage of the Monte Carlo methods  will become even
greater for larger sample sizes.   Note that in
cases such as catalytic activity, sensor response, or ligand
specificity,
 the experimental figure of merit would likely be exponential
in the values shown in Figure \ref{fig:results}, so that the success of the
Monte Carlo methods would be even more dramatic. A better calibration of
the parameters in Eq.\ (\ref{rpvm}) may be possible as more data becomes
available in the literature.

\subsection{Aspects of Further Development}
The space of composition and non-composition variables to search
in materials discovery experiments can be forbiddingly large.
Yet, by using Monte Carlo methods one can achieve an
effective search with a limited number of experimental samples.

Efficient implementations of the Monte Carlo search strategies
are feasible with
existing library creation technology.  Moreover
``closing the loop'' between library design and redesign
is achievable with the same database technology currently used
to track and record the data from  combinatorial chemistry experiments.
These multiple-round protocols, when combined with appropriate robotic
automation, should allow the practical
application of combinatorial chemistry to more complex and interesting
systems.

Many details need to be worked out in order
to flesh out the proposed protocols
for materials discovery. For example:
\begin{enumerate}
\item How rough are real
      figures of merit, and  can the Random Phase Volume Model
     be calibrated better?  
\item Can more of the 
      hierarchical structure of the variables be identified?
\item What are the best methods of manipulating the
     variables in the Monte Carlo?  
\end{enumerate}
Additional questions,
some of which this chapter has begun to answer,
include
how does the proximity to the global optimum scale with the
     number of samples and with the algorithm by which
     they are selected?
What is the best set of samples to choose for an optimal result,
     chosen ``all at once'' or in stages or sequentially?
What is the minimum number of samples required to make a 
     Monte-Carlo-based algorithm attractive as the driver?

\section{Protein Molecular Evolution}
The space to be searched in protein combinatorial chemistry experiments 
is extremely large.  Consider, for example, that a relatively
short 100 amino acid protein domain were to be evolved.
The number of possible amino acid sequences of this length is
$20^{100} \approx 10^{130}$, since there are 20 naturally occurring amino
acid residues.  Clearly, all of these sequences cannot be synthesized
and then screened for figure of merit in the laboratory.
Some means must be found for
 searching this space with the $10^4$ or so proteins  than can be
screened per day experimentally.

A hierarchical decomposition of the protein space can provide an
effective searching procedure.  It is known from protein structural
biology that proteins are encoded by DNA sequences, DNA sequences code
for amino acids, amino acids arrange into secondary structures, 
secondary structures arrange into domains, domains group to form
protein monomers, and protein monomers aggregate to form 
multi-protein complexes.  By sampling on each level of this hierarchy,
one is able to search the sequence space  much more effectively.
In this chapter, search strategies making use of the
DNA, amino acid, and secondary structure hierarchy will be described.
With this approach, functional protein space has a large,
yet manageable, number of dimensions. That is, in a
100 amino acid protein domain there
are approximate 10 secondary structures of 5 types (helices, loops,
strands, turns, and others) roughly yielding the
potential for $\approx10^7$ basic protein folds. 
Organization into secondary structural classes represents a dramatic
reduction in the complexity of sequence space, since
there are $\approx10^{170}$ different DNA sequences and
$\approx10^{130}$ different amino acid sequences in this space.

Sampling on the different levels of protein structure is
analogous to combination of different move
types in a Monte Carlo simulation.
A variety of moves, from small, local moves to large, global moves,
are often incorporated in the most successful Monte Carlo simulations.
While protein molecular evolution is carried out in the laboratory,
and Monte Carlo simulations are carried out \emph{in silico}, the parallels
are striking.  One of the most powerful new concepts in Monte Carlo is
the idea that moves should be ``biased'' \cite{Frenkel_book}. That is, small
moves, such as the Metropolis method, sample configuration space
rather slowly.  Larger moves are preferred, since they sample 
the space more rapidly.  Large moves are usually rejected, however,
since they often lead the system into a region of high energy.
So that the large moves will be more successful, a bias towards
regions that look promising is included.  Such biased Monte Carlo
simulations have been a factor of $10^5$ to $10^{10}$ 
times more efficient than
previous methods, and they have allowed the examination of systems previously
uncharacterizable by molecular simulation 
techniques \cite{Frenkel,SmitV,dePablo,review_83}.

In this section, the possibility of evolving protein molecules
by strategies similar to biased Monte Carlo will be explored.
The large moves of Monte Carlo are implemented by changing an
evolving protein at the secondary structure level.  These evolutionary
events will be biased, in that the amino acid sequences inserted will
be chosen so that they code for viable secondary structures.
The concept of bias also applies at the amino acid level, where
different DNA sequences coding for the same amino acids
can lead to different propensities for future evolution.

\subsection{What is Protein Molecular Evolution}

Protein molecular evolution can be viewed as combinatorial chemistry
of proteins.  Since protein sequence space is so large, most 
experiments to date have sought to search only small regions.
A typical experiment seeks to optimize the
figure of merit of an existing protein.  For example, an improvement in 
the selectivity or activity of an enzyme might be sought.  Alternatively,
an expansion in the operating range of an enzyme might be sought 
to higher temperatures or pressures.
This improvement would be achieved by changing, or evolving, the amino
acid sequence of the enzyme.

A more ambitious goal would be the \emph{ab initio}
evolution of a protein with a specific function.  That is,
nothing would be known about the desired molecule, except that
it should be a protein and that
an experimental screen for the desired figure of merit is available.
One might want, for example, an enzyme that catalyzes an unusual reaction.
Or one might want a protein that binds a specific substrate.
Or one might want a protein with an unusual fluorescence spectrum.
The \emph{ab initio} evolution of a protein has never been
accomplished before.  Such a feat would be remarkable.  Natural
biological diversity has evolved despite the essentially infinite
complexity of protein sequence.  Replication of this feat in the
laboratory would represent substantial progress, and
mimicking this feat of Nature is a current goal in the molecular
evolution field.  Indeed, the protocols described in this section
are crafted with just this task in mind.

A still more ambitious goal would be the evolution of a multi-protein 
complex.  This is a rather challenging task, due to the increased
complexity of the space to be searched.  The task can be made
manageable by asking a rather general evolutionary question.
One can seek to evolve, for example, a multi-protein complex that
can serve as the coat protein complex for a virus.  Since there are many
proteins that may accomplish this task, this evolutionary task may not
be as specific and difficult as it might seem at first.

The most ambitious goal for laboratory evolution that has been
imagined is the evolution of new life forms.  Evolution on this scale
requires changes not only at the secondary structure scale, but also
at the domain, protein, and protein pathway scale.  Due to their
simplicity, viruses or phage would be the most likely targets of such
large-scale evolution attempts.  It is unclear how 
new life forms would be distributed in terms of pathogenicity, and so
such experiments should be approached with caution.

The hierarchical decomposition of sequence space will allow effective
molecular evolution if there are many proteins with a high value of
any particular figure of merit.  That is, if only one out of
$10^{130}$ small protein domains exhibits a high score on
a particular figure of merit, this protein is 
unlikely to be identified.
 On the other hand, if many proteins score highly on the
figure of merit, only a subset of these
molecules need be sampled.  This same issue arises in conventional Monte Carlo
simulations.  Sampling all of configuration space
is never possible in  a simulation, 
yet ensemble averages and experimental behavior
can be reproduced by sampling representative configurations.
 That life on our planet
has evolved suggests that there is a great
redundancy in protein space \cite{len15},
and so one may hope to search this space 
experimentally with sufficiently powerful moves.

\subsection{Background on Experimental Molecular Evolution}
There are some constraints  on molecular evolution as it is carried 
out in the laboratory.  There are constraints arising from
limitations of molecular biology, \emph{i.e.}\ only certain types of
moves are possible on the DNA that codes for the protein.
There are also constraints arising from technical limitations,
\emph{i.e.}\ only a certain number of proteins can be screened for
figure of merit in a day.

Existing approaches to the evolution of general proteins are essentially
limited to changes at the single base level.  Somewhat more
sophisticated methods are  available for evolution of antibodies, but
this is a special case that will not be considered here.
  The first type of evolutionary change that is
possible in the laboratory is a base substitution. Base substitutions are
naturally made as DNA is copied or amplified by PCR.  The rate at which
base substitutions, or mistakes in the copying of the template DNA,
are made can be adjusted by varying experimental conditions, such as
the manganese and magnesium ion concentrations.  
It is important to note that these
base substitutions are made without knowledge of the DNA sequences of the
evolving proteins.   Equally important is that these base substitutions are
made without the use of a chemical synthesizer.  These changes are made
naturally within the context of efficient molecular biology methods.
Another means of modifying an existing protein is to use random or
directed mutagenesis to change specific DNA bases so that they code for
random or specified amino acids.  The approach requires both that the
DNA sequence be known and that it subsequently be synthesized by chemical
means.  Such a laborious  approach is not practical in high-throughput
evolution experiments, where typically $10^4$ proteins are simultaneously
modified and evolved per day.

Since the average length of a human gene is roughly 1800 bases, base-by-base
point mutation will achieve significant evolution only very slowly.
More significantly, the figure-of-merit landscape for protein function is
typically quite rugged.  Base mutation, therefore, invariably ceases
evolution at a local optima of the figure of merit.  Base mutation can
be viewed as an experimental method for local optimization of 
protein figures of merit.

Much of the current enthusiasm for protein molecular evolution is due to the
discovery of DNA shuffling by Pim Stemmer in 1994 
\cite{len7}.  DNA shuffling is a method for evolving an existing 
protein to achieve a higher figure of merit.
  The great genius of Stemmer was to develop a method
for combining beneficial base mutations that  is naturally
accomplished with the tools of molecular biology and that does not require
DNA sequencing or chemical synthesis.  The method is successful because
combination of 
base mutations that were individually beneficial 
is likely to lead to an even higher figure of merit than is achieved by
either mutation alone.  Of course, this will not always be true, but
the extent to which it is true is the extent to which DNA shuffling
will be an effective technique.  DNA shuffling, combined
with base mutation, is the current state of the art experimental
technique for protein molecular evolution.

Table \ref{table1} lists a few of the protein systems that have
been evolved by the Stemmer group.
\begin{table}[tbp]
\caption{Genes and operons evolved by DNA shuffling, from \cite{len11}.}
\label{table1}
\begin{center}
\begin{tabular}[t]{lccc}
\hline
\hline
System  & Improvement & Size & Mutations\\
\hline
TEM-1 $\beta$-lactamase & Enzyme activity    & 333 aa & 6 aa\\
                        & 32000-fold\\
$\beta$-galactosidase   & Fucosidase activity & 1333 aa & 6 aa\\
                        & 66-fold\\
Green fluorescence protein & Protein folding & 266 aa & 3 aa\\
                        & 45-fold\\
Antibody                & Avidity           & 233 aa & 34 aa$^1$\\
                        & $> 400$-fold\\
Antibody                & Expression level  & 233 aa & 5 aa\\
                        & $ 100$-fold\\
Arsenate operon         & Arsenate resistance & 766 aa & 3 aa\\
                        & 40-fold\\
Alkyl transferase       & DNA repair       & 166 aa & 7 aa\\
                        & 10-fold\\
Benzyl esterase         & Antibiotic deprotection & 500 aa & 8 aa\\
                        & 150-fold\\
tRNA synthetase         & Charging of & 666 & not \\
                        & engineered tRNA && determined\\
                        & 180-fold\\
\hline
\hline
\end{tabular}
\end{center}
\hbox{}$^1$This was a case of family shuffling \cite{len8},
so most of these changes were between homologous amino acids.
\end{table}
Evidently, DNA shuffling is highly effective at improving the 
function of an existing protein, much more effective than is simple
base mutation.  The specificity of an enzyme can
even be altered, as in the conversion of a
$\beta$-galactosidase into a fucosidase.
A rough median of the improvement factors is about 100.
Most important, however, is that all of these improvements were achieved
with a relatively small number of amino acid changes. 
On average, only 6 amino acids were altered out of roughly 400 total
residues in the protein.  As with base mutation, then, DNA shuffling
is able to search sequence space only locally.  After a small number of
amino acid changes, DNA shuffling produces a protein with a 
locally rather than globally optimal figure of merit.

The current state of the art experimental techniques for protein
molecular evolution can be viewed as local optimization procedures
in protein sequence space.
Alternatively, they can be viewed as experimental implementations of
 simple, or Metropolis, Monte Carlo procedure.  
By using  our intuition regarding the design of powerful, biased Monte Carlo
algorithms, we can develop more powerful experimental protocols for
molecular evolution.

Interestingly, theoretical treatments of evolution, whether in Nature
or in the laboratory, tend to consider only the effects of point mutation
\cite{len15,Volkenstein}.  Indeed, interesting theories regarding
the evolutionary potential of point mutations have been developed.
As shown experimentally, however, point mutation is incapable of
significantly evolving proteins  at substantial rate.  Even the more
powerful technique of DNA shuffling searches protein space merely locally.  
Only with the inclusion of more dramatic moves, such as changes at the
level of secondary structures,  can
protein space be searched more thoroughly.

\subsection{The Generalized NK Model}
In order to validate the molecular evolution protocols to 
be  presented, a model that relates amino acid sequence to protein
function is needed.  Of course, the real test of these protocols 
should be experimental, and I hope that these experiments will be
forthcoming. In order to stimulate interest in the proposed
protocols, their effectiveness will be simulated on a model of
protein function. 
Such a model would seem to be difficult to construct.
It is extremely difficult to determine the three-dimensional structure
of a protein given the amino acid sequence.  Moreover, it is
extremely difficult to calculate any of the typical figures of merit
given the three-dimensional structure of a protein.

It is fortunate that a model that relates figure of merit
to amino acid sequence for a specific protein is not needed.
The requirement is simply a model
that produces figure-of-merit landscapes in sequence space that
are analogous to those that would be measured in the laboratory on an
ensemble of proteins.  This type of model is easier to construct,
and a  random energy model can be used to accomplish the task.

The generalized NK model is just such a random 
energy model.  The NK model was first introduced
in order to model combinatorial
chemistry experiments on peptides \cite{len14,len15,len16}.  It was
subsequently generalized to account for secondary structure in
real proteins \cite{len17}.
 The model was further generalized
to account for interactions between the secondary structures and
for the presence of a binding pocket \cite{Deem3}.

This generalized NK model assigns a
unique figure of merit to each evolving protein sequence. This model,
while a simplified description of real proteins, captures much of the
thermodynamics of protein folding and ligand binding.
The model takes into account the formation of
secondary structures via the interactions of amino acid side chains
 as well as the interactions between secondary structures
within proteins. In addition, for specificity, 
the figure of merit is assumed to be a binding constant, and so
the model includes a contribution
representing binding to a substrate.
 The combined
ability to fold and bind substrate is what will be
optimized or evolved.
That is, the direction of the protein evolution will be based upon
the figure of merit returned by this generalized NK model.
This generalized NK model contains several parameters, and a reasonable
determination of these parameters is what allows the model to compare
successfully with experiment.

The specific energy function used as the selection criterion in the
molecular simulations is
\beq{101}
 U = \sum_{\alpha = 1}^M U_\alpha^\mathrm{sd} +
\sum_{\alpha > \gamma = 1}^M U_{\alpha \gamma}^\mathrm{sd-sd}
+ \sum_{i = 1}^P U_i^\mathrm{c} \ .
\eeq
This energy function is composed of three parts: secondary structural
subdomain energies ($U^\mathrm{sd}$), subdomain-subdomain interaction energies
($U^\mathrm{sd-sd}$), and chemical binding energies ($U^\mathrm{c}$).  Each of these
three energy terms is weighted equally, and each has a magnitude near
unity for a random sequence of amino acids. In this NK based
simulation, each different type of amino acid behaves as a completely
different chemical entity; therefore, only $Q=5$ five chemically distinct
amino classes are considered (\emph{e.g.}, negative, positive, polar,
hydrophobic, and other). Interestingly, restricted alphabets of
amino acids not only are capable of producing functional proteins 
\cite{len19,len20} 
but also
may have been used in the primitive genetic code
 \cite{len21,len22}.
The evolving protein will be a relatively short 100 amino acid protein domain.
Within this domain will be roughly $M=10$ secondary structural subdomains,
each $N=10$
amino acids in length. The subdomains belong to one of $L=5$ different types
(\emph{e.g.}, helices, strands, loops, turns, and others). This gives
L different ($U^\mathrm{sd}$) energy functions of the NK form
\cite{len14,len15,len16,len17}. 
\beq{102}
U_\alpha^\mathrm{sd} = \frac{1}{\left[ M(N-K)\right]^{1/2}}
\sum_{j = 1}^{N-K+1} \sigma_\alpha \left(
a_j, a_{j+1}, \ldots, a_{j+K-1}
\right) \ .
\eeq
The degree of complexity in the interactions between the amino acids
is parameterized by the value of $K$.  Low values of $K$ lead to 
figure-of-merit landscapes upon which evolution is easy, and
high values of $K$ lead to extremely rugged landscapes upon which
evolution is difficult.  Combinatorial chemistry experiments on 
peptides have suggested the value of $K=4$ as a reasonable
one \cite{len16}. Note that the definition of $K$ here is
one greater than the convention in \cite{len14,len15,len16}.
The quenched, unit-normal random        
number $\sigma_\alpha$ in Eq.\ \ref{102} is different for each
value of its argument for each of the $L$ classes. This random form
mimics the complicated amino acid side chain interactions
within a given secondary structure. The energy of interaction
between secondary structures is given by
\bey{103}
U_{\alpha \gamma}^\mathrm{sd-sd} &=& \left[
\frac{2}{D M (M-1)} \right]^{1/2} \nonumber \\
&&\times 
\sum_{i=1}^D \sigma_{\alpha \gamma}^{(i)}
\left(
a_{j_1}^\alpha, \dots,
a_{j_{K/2}}^\alpha;
a_{j_{K/2+1}}^\gamma, \ldots
a_{j_{K}}^\gamma
 \right) \ .
\eey
The number of interactions between secondary structures is set at 
$D=6$. Here the unit-normal weight, $\sigma_{\alpha \gamma}^{(i)}$,
and the interacting amino acids, $\{j_1,\ldots,j_K\}$, are selected at
random for each interaction $(i, \alpha, \gamma)$. The chemical
binding energy of each amino acid is given by
\beq{104}
U_i^\mathrm{c} = \frac{1}{\sqrt P} \sigma_i \left( a_i \right) \ .
\eeq
The contributing amino acid, $i$, and the unit-normal weight of the
binding, $\sigma_i$, are chosen at random.  
A typical binding pocket is composed of five
amino acids, and so the choice of
$P=5$ is made.

\subsection{Experimental Conditions and Constraints}
A typical protein evolution experiment starts with an initial protein
sequence.  This sequence is then copied to a large number of identical
sequences.  All of these sequences are evolved, or mutated, in 
parallel.  After one round of mutation events, the proteins are screened for
figure of merit.  This screening step is typically the rate limiting
step, and so the efficiency of this step determines how many proteins can be
evolved  in parallel.  For typical figures of merit, 10000 proteins
can be screened in a day.  If selection,
that is use of a screen based upon whether an organism lives or dies, 
were performed instead, 
$10^9-10^{15}$ proteins could be screened in a day.  Selection is a special
case, however, and so the more conservative case of screening
10000 proteins per day is considered.

After the screen, the proteins are ranked according to their
measured value of the figure of merit.  
Typically, the top 
$x$ percent of the sequences are kept for the next round of mutation.
The parameter $x$ is to be adjusted experimentally.  In the simulated
evolutions, the value of $x = 10\%$ was always found to be optimal.
Other methods for selecting the proteins to keep for the next round
have been considered.  For example, keeping proteins proportional
to $\exp(-\beta U)$ has been considered.
This strategy seems to work less well than the top $x$ percent method.
The main reason seems to be that in the top $x$ percent method,
the criterion for selecting which sequences to keep adjusts naturally
with the range of figures of merit found in the evolving sequences.
After the top $x$ percent sequences are selected, they are copied back
up to a total of 10000 sequences.  These sequences are the input for
the next round of mutation and selection.

In the simulated molecular evolutions, 
the experiment is continued for 100 rounds.
This is a relatively large number of rounds to carry out experimentally.
With the most powerful protocols, however, it is possible to evolve proteins
\emph{ab initio}.  This feat has not been achieved to date in the laboratory.
In order to mimic this feat of Nature, one should be willing to do
some number of rounds.

\subsection{Several Hierarchical Evolution Protocols}

\subsubsection{Amino acid substitution}
\begin{figure}[tbp]
\centering
\leavevmode
\psfig{file=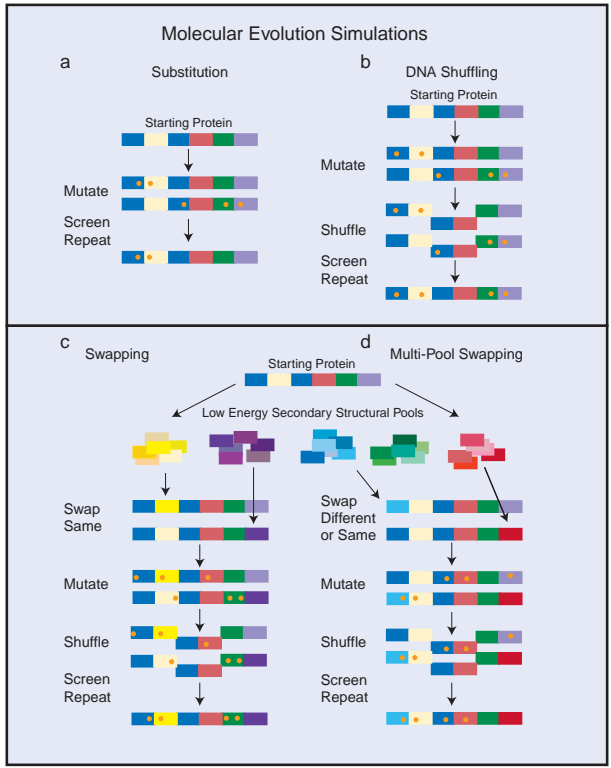,height=4.0in}
\caption{Schematic diagram of the simulated molecular evolution
protocols, from \cite{Deem3}.
 a) Simulation of molecular evolution via base substitution
(substitutions are represented by orange dots). b) Simulated DNA
shuffling showing the optimal fragmentation length of 2 subdomains. c)
The hierarchical optimization of local space searching: The 250
different sequences in each of the 5 pools (\emph{e.g.}, helices, strands,
turns, loops, and others) are schematically represented by different
shades of the same color. d) The multi-pool swapping model for
searching vast regions of tertiary fold space is essentially the same
as in the Figure \ref{fig:protocol}c except that now
sequences from all 5 different structural pools can be swapped into
any subdomain.
Multi-pool swapping
allows for the formation of new tertiary structures by changing the
type of secondary structure at any position along the protein.
}
\label{fig:protocol}
\end{figure}

To obtain a base line for searching fold space, molecular
evolution is first simulated
via simple mutagenesis (see Figure \ref{fig:protocol}a).  Simulated
evolutions by amino acid substitution lead to significantly improved
protein energies, as shown in Table \ref{table2}.
 These evolutions always terminated at local energy
minima, however.  This trapping is due to
the difficulty of combining the large number of
correlated substitutions necessary to generate new
protein folds.  Increasing the screening stringency in later rounds
did not improve the binding constants of simulated proteins, most
likely due to the lack of additional selection criteria such as
growth rates. Although only non-conservative mutations were directly simulated,
conservative and synonymous neutral mutations are not excluded and
can be taken into account in a more detailed treatment.
Indeed, the
optimized average mutation rate of 1 amino acid
substitution/sequence/round is equivalent to roughly 1-6 random base
substitutions/round.
\begin{table}[tbp]
\caption{Results of Monte Carlo simulation of the evolution protocols,
from \cite{Deem3}.$^1$}
\label{table2}
\begin{center}
\begin{tabular}[t]{cccc}
\hline
\hline
Evolution Method &  $U_{\rm start}$&   
$U_{\rm evolved}$&    $k_{\rm binding}$\\
\hline
Amino Acid Substitution &     -17.00 &        -23.18 &             1\\
DNA Shuffling        &   -17.00      &   -23.83      &        100\\
Swapping            &0       & -24.52             & $1.47 \times 10^4$\\
Mixing              &0       & -24.88           &   $1.81 \times 10^5$\\
Multi-Pool Swapping$^2$ &      0 &       -25.40 &    $8.80 \times 10^6$\\
\hline
\hline
\end{tabular}
\end{center}
\hbox{}$^1$The starting polypeptide energy of -17.00 comes from
a protein-like sequence (minimized $U^\mathrm{sd}$), and 0 comes from a random
initial sequence of amino acids. The evolved energies and binding constants
are median values. The binding constants are calculated 
from $k_{\rm binding} = a e^{-b U}$,
where $a$ and $b$ are constants determined by normalizing the binding
constants achieved by point mutation and shuffling to 1 and 100,
 respectively.\\
\hbox{}$^2$Note that the energies
and binding constants achieved via multi-pool swapping represents
typical best evolved protein folds.
\end{table}

\subsubsection{DNA shuffling}
DNA shuffling improves the search of local fold space via a random yet
correlated combination of homologous coding fragments that contain limited
numbers of beneficial amino acid substitutions.  As in experimental
evolutions \cite{len7,len8,len9,len10}, the simulated shuffling
improved protein function significantly better than did point
mutation alone (see Table \ref{table2} and Figure \ref{fig:protocol}b). However, local
barriers in the energy function 
also limit molecular evolution via DNA shuffling.
For example,  when the screen size was increased from 10000 to 
20000 proteins per round, no
further improvement  in the final evolved energies was seen.
Interestingly, the
optimal simulated DNA shuffling length of 20 amino acids (60 bases) is
nearly identical to fragment lengths used in experimental protocols
\cite{len8}.

\subsubsection{Single-pool swapping}
In Nature, local protein space can be rapidly searched by the directed
recombination of encoded domains from multi-gene pools.
A prominent example is the creation of the primary antibody
repertoire in an
adaptive immune system. These events are generalized by
simulating the swapping of amino acid fragments from 5 different
structural pools representing helices, strands, loops, turns, and
others (see Figure \ref{fig:protocol}c).  During the swapping step, subdomains were
randomly replaced with members of the same secondary structural pools
with an optimal probability of 0.01/subdomain/round.
The simulated evolution  of
the primary fold is limited by maintaining the linear
order of swapped secondary structure types. The addition of the
swapping move was so powerful that it was possible to achieve binding
constants 2 orders of magnitude higher than in shuffling simulations
(see Table \ref{table2}).  Significantly, these improved binding
constants were achieved starting with
10-20 times less minimized structural subdomain material.

\subsubsection{Mixing}
\begin{figure}[tbp]
\centering
\leavevmode
\psfig{file=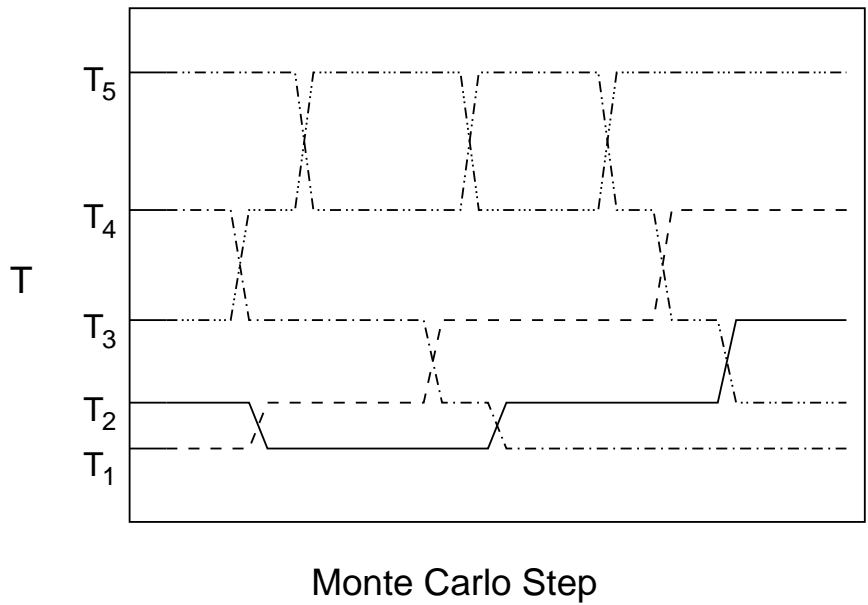,width=4.5in}
\caption{A schematic drawing of the swapping taking place during a
parallel tempering simulation.}
\label{fig:temper}
\end{figure}

Parallel tempering is a powerful statistical method that often allows
a system to escape local energy minima \cite{Geyer}.
This method simultaneously simulates several 
systems at different temperatures, allowing systems
at adjacent temperatures to swap configurations.
The swapping between high- and low-temperature systems allows for
an effective searching of configuration space.
This method achieves rigorously correct canonical sampling, and it
significantly reduces the equilibration time in a simulation. Instead
of a single system, 
a larger ensemble with $n$ systems is considered in parallel tempering,
and each system is
equilibrated at a distinct temperature $T_i$, $i=1,\ \ldots,\ n$.
The procedure in parallel tempering is illustrated in 
Figure \ref{fig:temper}

 The
system with the lowest temperature is the one of our interest; the
higher temperature systems are added to aid in the equilibration of
the system of interest. In
addition to the normal Monte Carlo moves performed in each system,
swapping moves are proposed that exchange the configurations between
two systems $i$ and $j=i+1$, $1\leq i< n$.
The higher
temperature systems are included solely to help the lowest temperature
system to escape from local energy minima via the swapping moves.  To
achieve efficient sampling, the highest temperature should be such
that no significant free energy barriers are observed. So that the
swapping moves are accepted with a reasonable probability, the energy
histograms of systems adjacent in the temperature ladder should
overlap.

  In Nature, as
well, it is known that genes, gene fragments, and gene operons are
transfered between species of different evolutionary complexity (\emph{i.e.},
at different ``temperatures'').   By analogy, 
limited population mixing is performed
among several parallel swapping experiments
by randomly exchanging evolving proteins at an optimal
probability of 0.001/protein/round. These mixing simulations optimized local
space searching and
achieved binding constants $\approx10^5$
higher than did base substitution alone (see Table \ref{table2}). Improved
function is due, in part, to the increased number of events in parallel
experiments.  Indeed, mixing may occur in Nature when
the evolutionary target function changes with time.
That is, in a dynamic environment
with multiple selective pressures, mixing would 
be especially effective when the rate of evolution of an isolated
population is slower than the rate of environmental change. 
It has also been argued that spatial heterogeneity in drug
concentration, a form of ``spatial parallel tempering,'' facilitates the
evolution of drug resistance \cite{Perelson}.

\subsubsection{Multipool swapping}
\begin{figure}[tbp]
\centering
\leavevmode
\psfig{file=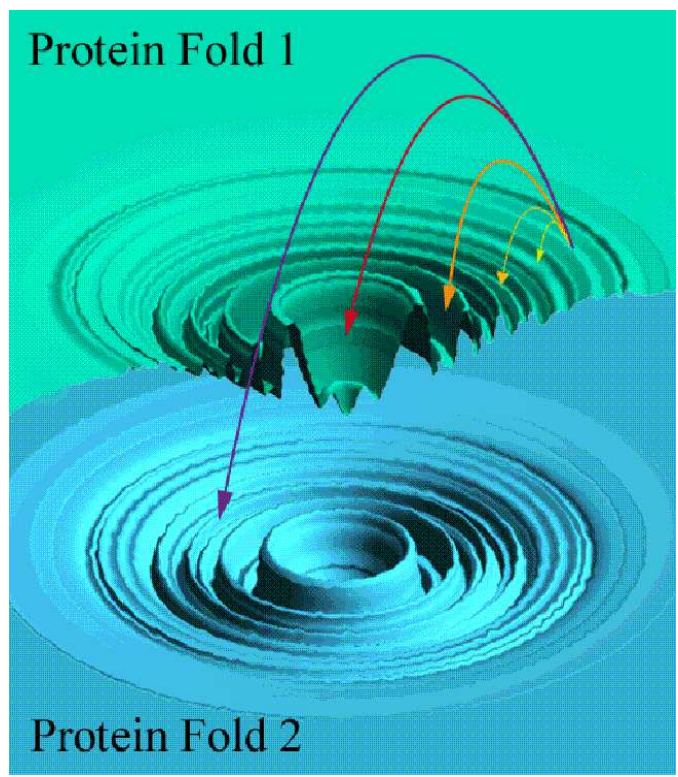,height=4.0in}
\caption{
Schematic diagram representing a portion of the
high-dimensional protein composition space,
from \cite{Deem3}.  The three-dimensional energy
landscape of Protein Fold 1 is shown in cut away. The arcs
with arrowheads represent the ability of a given molecular evolution
process to change the composition and so to
traverse the increasingly large barriers in the energy function. The smallest
arc represents the ability to evolve improved fold
function via point mutation. Then in increasing order: DNA shuffling,
swapping, and mixing. Finally, the
multi-pool swapping protocol allows an evolving system to move
to a different energy landscape representing a new tertiary fold
(bottom).
}
\label{fig:fold}
\end{figure}

The effective navigation of protein space requires the discovery and
selection of tertiary structures. To model the large scale search of
this space, a random polypeptide
sequences was used as a starting point, and 
the swapping protocol was repeated. Now, however, secondary
structures from all 5 different pools were permitted
 to swap in at every subdomain
(see Figure \ref{fig:protocol}d).  This multi-pool swapping approach 
evolved proteins with binding constants $\approx10^7$ better than did
amino acid substitution of a protein-like starting sequence (see Table
\ref{table2}).  This evolution was accomplished by the
random yet correlated
juxtaposition of different types of low energy secondary structures.
 This approach dramatically improved specific
ligand binding while efficiently discovering new tertiary structures
(see Figure \ref{fig:fold}).  Optimization of the rate of these hierarchical molecular
evolutionary moves, including relaxation of
the selection criteria, enabled the
protein to evolve despite the high rate of failure for these dramatic
swapping moves.
 Interestingly, of all the molecular evolutionary
processes modeled, only multi-pool swapping demonstrated chaotic
behavior in
repetitive simulations.  This chaotic behavior was likely due to
the discovery of different model folds that varied
in their inherent ability to serve as scaffolds for ligand specific binding.

\subsection{Possible Experimental Implementations}
An important motivation of this work was that the proposed protocols 
must be experimentally feasible.  Indeed, the ultimate test of the
effectiveness of these protocols will be experimental.
It is hoped that 
the search of large regions of protein space apparently possible
with these methods will identify new protein folds
and functions of great value to basic, industrial, and
medical research.

The main technical challenge posed by the swapping protocols is the
non-homologous recombination necessary to swap the 
DNA that codes for different secondary structures
into the evolving proteins.  One approach would be
to generate multiple libraries of synthetic oligonucleotide
pools 
\cite{len25,len26} 
encoding the different secondary 
subdomain structures. Asymmetric, complementary encoded linkers with
embedded restriction sites would make the assembly, shuffling, and
swapping steps possible.

Alternatively, the techniques of 
ITCHY \cite{Benkovic1} and SCRATCHY 
\cite{Benkovic2} may be used to accomplish the non-homologous swapping of
secondary structures required within our protocol.

Finally, exon shuffling may be used to perform the non-homologous
swapping events.  In this case, the pools of secondary structures
would be encoded with exons of a living organism, such
as \emph{E. coli}.  There is precedent for such use of
exon shuffling, both at the DNA \cite{len24} and RNA 
\cite{Moran} level.

\subsection{Life has Evolved to Evolve}
Although the focus has been on the higher levels of the evolution
hierarchy, because that is where the biggest theoretical and
experimental gap lies, all levels are important.  In particular, 
the details of how point mutation assists protein evolution
is important.

\begin{figure}[tbp]
\centering
\leavevmode
\psfig{file=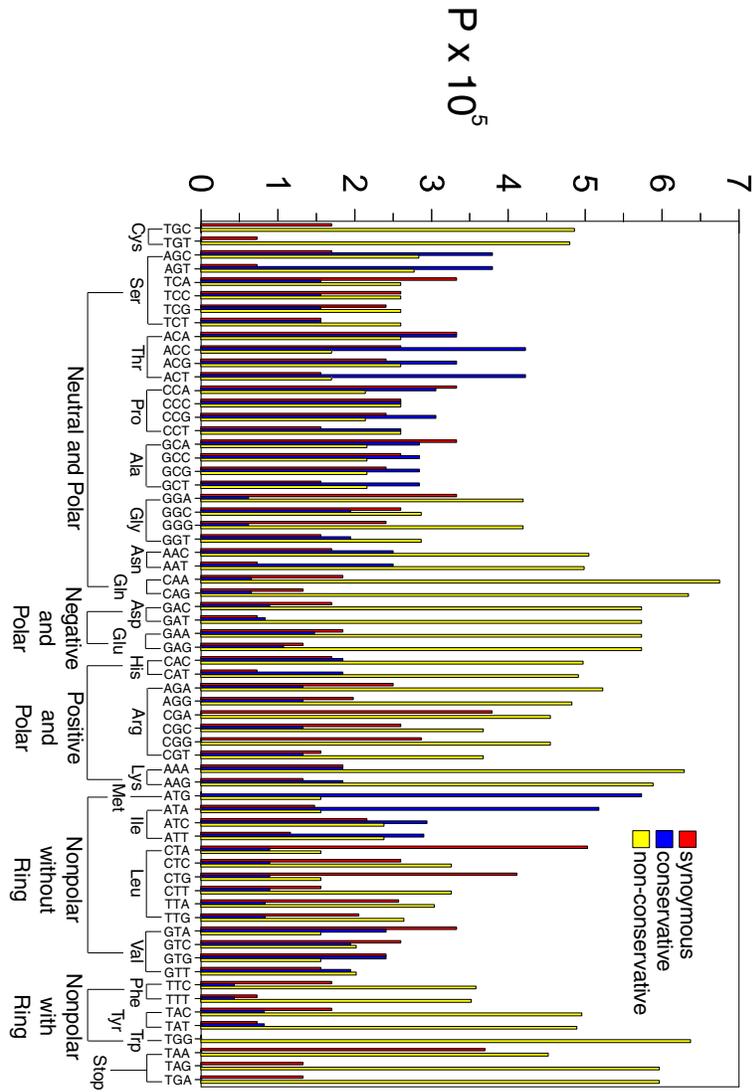,height=6.0in,angle=-90}
\caption{
Shown are the probabilities of a given codon mutating
in a synonymous (unfilled), conservative (hatched), and
non-conservative (filled) way in one round of replication under
error prone conditions.  The codons are grouped by the amino acid
encoded, and the amino acids are group by category.
}
\label{fig:mutate}
\end{figure}

DNA base mutation leads only indirectly to changes in protein
expression.  How mutations occur in the bases and how these mutations
lead to codon changes, and so amino acid changes, is not purely
random.  The inherent properties of the genetic code and biases in
the mechanisms of DNA base substitution are perfectly suited for the 
``neutral'' search of local space.  Previously, the genetic code has
been presented as a nodal or hypercubic structure to illustrate these
relationships \cite{len31,Montano}.  It seems
preferable to view the standard
genetic code quantitatively as a $64 \times 64$ two-dimensional
matrix. 
Seeds of this approach can be found in
Kepler's work regarding
evolvability in immunoglobulin genes \cite{Kepler,Kepler2,nsf5_7}.
The values in this matrix are the probabilities of a
specific codon mutating to another by a single base change under
error-prone conditions, \emph{e.g.}
mutator strains of bacteria, error prone PCR, or somatic
hypermutation.  Assuming each base mutates independently
in the codon, this matrix can be calculated from a simpler $4 \times 4$
matrix of base mutation probabilities.  The base mutation matrix
can be extracted from available experimental data \cite{Smith96}.
A synonymous transition probability can be defined for each
codon, which is the probability per replication
 of a base change that leads to a
codon that codes for the same amino acid.  
A conservative transition probability can further be defined, which is the
probability per replication that a base change leads to a conservative 
mutation.  Finally, a non-conservative 
transition probability can be defined, which is
the probability per replication that a base change leads to a non-conservative
mutation.    The conservative and non-conservative mutation
probabilities can be viewed as defining the evolutionary potential
of each codon: codons with high conservative and non-conservative
mutation rates can be said to exhibit a high evolutionary potential.
These mutation tendencies are shown in Figure \ref{fig:mutate}.
In general, amino acids that exhibit a dramatic functional property, such as
the charged residues, the ringed residues, cysteine, and tryptophan, 
tend to mutate at higher non-conservative rates that allows for the
possible deletion of 
the property.  Amino acids that are more generic, such as
the polar neutral residues and the non-polar non-ringed residues,
tend to mutate at higher conservative rates that allows
sequence space to be searched for similar favorable contacts.

There is a connection between condon usage and DNA shuffling.
In the most successful DNA shuffling experiments by the
Stemmer group, codon assignments of the initial coding sequences are
optimized for expression.  This assignment
typically increases the non-synonymous mutation rates
described above.    In particular, this assignment tends to increase the
conservative mutation rate for the generic amino acids and the
non-conservative mutation rate for the dramatically-functional
amino acids.
Codon usage, then, has already implicitly
been used to manipulate mutation rates. Explicit consideration of
the importance of codon assignment would
be an interesting amino-acid level refinement of existing molecular
evolution protocols.  Optimized protocol parameters can
be identified, taking into account the detailed codon usage information.
Similarly, the codon potential matrix can be used
in the design of the pools of secondary structures in the
swapping-type molecular evolution protocols.
That is, DNA can be chosen that 
codes for the secondary structures that
i) tends not to mutate, ii) tends to mutate to synonymous sequences,
iii) tends to mutate to conservative sequences,
or iv) tends to mutate to non-conservative sequences.

In Nature there are numerous examples of exploiting codon potentials
in ongoing evolutionary processes \cite{nsf5_7}.  In the V regions of
encoded antibodies, high-potential serine codons such as AGC
are found predominately in the encoded CDR loops while the encoded
frameworks contain low-potential serine codons such as TCT.
Unfortunately, antibodies and drugs are often no match for the
hydrophilic, high-potential codons of ``error-prone'' pathogens.
The dramatic mutability of the HIV gp120 coat protein is one
such example.  One can envision a
scheme for using codon potentials to target disease epitopes that
mutate rarely (\emph{i.e.}, low-potential) and unproductively 
(\emph{i.e.}, become stop, low-potential, or structure-breaking
codons).  Such a therapeutic scheme is quite simple, and so could be quite
generally useful against diseases that otherwise tend to become
drug resistant.

\subsection{Natural Analogs of these Protocols}

During the course of any evolutionary process, proteins become trapped
in local energy
minima. Dramatic moves, such as swaps and juxtaposition, are needed to
break out of these
regions. Dramatic moves are usually deleterious, however. The evolutionary
success of these
events depends on population size, generation time, mutation rate,
population mixing,
selective pressure or freedom, such as successful genome
duplications or the
establishment of set-aside cells \cite{len27}, and the mechanisms that transfer
low energy, encoded structural domains.

By using the analogy with Monte Carlo to design ``biased'' moves
for molecular evolution, a swapping-type move has been derived
that is similar to several mechanisms of natural evolution.
Viruses and transposons, for example, have
evolved large-scale integration mechanisms \cite{len13}.
Exon shuffling is also a generator of diversity, and 
a possible scenario is that
 exon shuffling generated the primordial fold diversity
\cite{len28,len29,len30}. 
Alternatively, random swapping by horizontal transfer \cite{len12},
rearrangement,
recombination, deletion, and insertion can lead to high
in-frame success rates in genomes with high densities of
coding domains and reading frames, as in certain prokaryotes and
mitochondria.

While inter- and intra-species exchange of DNA is often thought to
occur primarily on the scale of genes and operons, shorter exchange
often occurs.  Indeed, the most prevalent exchange length within
\emph{E. coli} is on the order of several hundred to a thousand base pairs
\cite{nsf5_20}.
Similarly, analysis of the evolution of vertebrate cytochrome-c
suggests that transfer of segments significantly smaller than a
single gene must have occurred \cite{nsf5_20}.

Indeed, swapping mutations leading to significant diversity are not
rare in Nature.  \emph{Neisseria meningitidis} is a frequent cause of
meningitis in sub-Saharan Africa \cite{nsf5_3}.  The Opa proteins are
a family of proteins that make up part of the outer coat of this
bacteria.  These proteins undergo some of the same class switching
and hypervariable mutations as antibody domains.  A significant
source of diversity also appears to have come from interspecies transfer
with \emph{Neisseria gonorrhoeae} \cite{nsf5_3}.  Both the surface
coat proteins and pilon proteins of \emph{N. gonorrhoeae} undergo
significant homologous recombination to produce additional diversity.
It appears
generally true that the intra- and interspecies
 transfer of short segments of genes is
common in \emph{E. coli}, \emph{Streptococci}, and
\emph{Neisseria}.
Rapid evolution of diversity such as this obviously poses a
significant challenge for therapeutic protocols.

Three dramatic examples of use of swapping by Nature are particularly
notable.   The first is the development of antibiotic resistance.  It
was originally thought that no bacteria would become resistant to
penicillin due to the many point mutations required for resistance.
Resistance occurred, however, within several years.  It is now known that
this resistance occurred through the swapping of pieces of DNA between
evolving bacteria \cite{Shapiro,Shapiro2}.
  One mechanism of antibiotic resistance was incorporation of
genes coding for $\beta$-lactamases. These genes, which directly
degrade $\beta$-lactam antibiotics, appear to be relatively ancient, and
their incorporation is a relatively simple example of a swapping-type
event.  These $\beta$-lactamases have continued
to evolve, however, in the presence of
antibiotic pressure, both by point mutation and by 
shuffling of protein domains via exon shuffling
\cite{Maiden,Medeiros}.
The bacterial targets of penicillin, the penicillin-binding 
proteins, are modular proteins that have undergone significant 
structural evolution since the introduction of penicillin
\cite{Massova,Goffin}.
This evolution was of a 
domain-shuffling form, and it is a more sophisticated example
of a Natural swapping-type move.
Multi-drug resistance is, of course, now a
major, current health care problem.
 The creation
of the primary antibody repertoire in vertebrates  is another example
of DNA swapping (of genes, gene segments, or pseudo-genes).
Indeed, the entire immune system mostly likely evolved from a single
transposon insertion some 450 million years ago \cite{nsf5_13,nsf5_15}.
This insertion, combined with
duplication and subsequent mutation of
a single membrane spanning protein, mostly likely lead to the
class switching apparatus of the primary repertoire.
  Finally, the evolution of \emph{E. coli} from
\emph{Salmonella} occurred exclusively by DNA swapping
\cite{len12}.  None of the phenotypic differences between these two
species is due to point mutation. Moreover, even the observed
rate of evolution due to DNA swapping, 31000 bases/million years,
is higher than that due to point mutation, 22000 bases/million years.
Even though a DNA swapping event is less likely to be tolerated
than is a point mutation, the more dramatic nature of the swapping event
leads to a higher overall rate of evolution.  This is exactly the
behavior observed in the simulated molecular evolutions.

\subsection{Concluding Remarks on Molecular Evolution}

DNA base substitution, in the context of the genetic code, is
ideally suited for the
generation, diversification, and optimization of local protein space
\cite{len21,len31}. However, the
difficulty of making the transition from one
productive tertiary fold to another limits
evolution via base
substitution and homologous recombination alone.
 Non-homologous DNA recombination, rearrangement, and
 insertion allow for
the combinatorial creation of productive tertiary folds via the novel
combination of suitable
structures. Indeed, efficient search of the high-dimensional
fold space requires a
hierarchical range of mutation events.

This section has
addressed from a theoretical point of view the question
of how protein space can be searched efficiently and thoroughly,
either in the laboratory or in Nature.   It was shown
that point mutation alone is incapable of evolving systems with
substantially new protein folds. It was further demonstrated that
even the DNA shuffling approach is incapable of evolving
substantially new protein folds.    The Monte Carlo simulations
demonstrated that non-homologous DNA ``swapping'' of low energy
structures is a key step in searching protein space.

More generally, the simulations demonstrated that the efficient search
of large regions of protein space requires a hierarchy of genetic
events, each encoding higher order structural substitutions.  It was
shown how the complex protein function landscape can be navigated with
these moves.  It was concluded that analogous moves have driven the
evolution of protein diversity found in Nature.  The proposed
moves, which appear to be experimentally feasible, would make an
interesting addition to the techniques of molecular biology.
An especially important application of the theoretical approach
to molecular evolution is
modeling the molecular evolution of disease.

There are many experimental applications of the technology for
 molecular evolution  \cite{len11}.
Perhaps some of the most significant are in the
field of human therapeutics.
    Molecular evolution can be used directly to
improve the performance of protein pharmaceuticals.
Molecular evolution can be used indirectly to 
evolve small molecule pharmaceuticals by evolving the pathways 
that code for small molecule synthesis in \emph{E. coli}.
Molecular evolution can be used for gene therapy and 
DNA vaccines.  Molecular evolution can be used to
produce recombinant protein vaccines or viral vaccines.
Finally, molecular evolution can be used to create modified
enzymatic assays in drug screening efforts.
The ability to develop new assays that do not infringe on
competitors' techniques is an important ability for
large pharmaceutical companies, given the current complex
state of patent claims.
There is a similar range of applications of molecular evolution in the
field of biotechnology.
As shown in Table \ref{table1}, many of the tools of
molecular biology can be improved or modified through the use of
molecular evolution. 

A wide variety of pest organisms and parasites, including
fungi, weeds, insects, protozoans, macroparasites, and
bacteria, have used evolutionary processes to evade chemical
control.  The range of evolutionary events exhibited by these
organisms is similar in spirit to the hierarchy of moves
present in the molecular evolution protocol (see Figure
\ref{fig:protocol}).  Bacteria provide one of the most pressing
examples of the problems posed by an evolving disease (a
``moving target'').  Although there undoubtedly have been
many selective pressures upon
bacteria, the novel pressure with the largest impact in the
last half century has been the worldwide use of antibiotics.
This background presence of antibiotics has lead to the
development of antibiotic resistance in many species of bacteria.
Indeed, multi-drug resistance is now a major health care issue, with
some strains resistant to all but one, or even all, known
antibiotics.

Interestingly, there is another strong pressure on evolving bacteria,
that of the vertebrate immune system.  This pressure is thought
to be responsible for mosaic, or modular as a result of
swapping-type events,
genes found in species of bacteria
not naturally genetically competent, such as
\emph{E. coli} and \emph{S. pyogenes} \cite{nsf5_11}.  In these cases,
the long-standing, strong selective pressure due to the interaction
with the immune system likely led to genetic exchange.

Kepler and Perelson have noted that a spatial heterogeneity
in the concentration of a drug can facilitate evolved resistance in
the disease organism \cite{Perelson}.  This occurs because regions
of low drug concentration provide a ``safe harbor'' for the disease,
where replication and mutation can occur.  The regions of high
disease concentration provide the selective pressure for the
evolution.  Explicit examples of this mode of evolution include the
role of spatial heterogeneity in the spread of insecticide resistance,
noncompliance to antibiotic regimes in the rise of resistance in the
tuberculosis bacterium, and heterogeneity within the body of the
protease inhibitor indinavir in the rise of resistant HIV-1 strains
\cite{Perelson}.
As noted above, this type of evolution is a spatial example of
parallel tempering, a technique that has proven to be very powerful
at sampling difficult molecular systems with many and large energy
barriers.  This analogy with parallel tempering suggests that
heterogeneities must be of great and intrinsic importance in natural
evolution.

 Qualitative changes in
protein space such as those modeled here allow viruses, parasites,
bacteria, and cancers to evade the immune system, vaccines,
antibiotics, and therapeutics.   All of these pathogens evolve,
to a greater or lesser degree, by large, swapping-type mutations.
The successful design of vaccines and
drugs must anticipate the evolutionary potential of both local and
large space searching by pathogens in response to therapeutic and
immune selection.  The addition of disease specific constraints to
simulations such as these  should be a promising approach for
predicting pathogen plasticity. 
Indeed, infectious agents will continue to evolve
unless we can force them down the road to extinction.

\section{Summary}
Significant opportunities exist for the 
application of ideas from statistical
mechanics to the burgeoning area of combinatorial chemistry.
While combinatorial chemistry was not invented by researchers in the
field of statistical mechanics, it is fair to say that perhaps it should
have been! 
The design of
effective experimental methods for searching composition space is
similar in concept to the design of effective Monte Carlo
methods for searching configuration space.
Optimization of the parameters in combinatorial chemistry
protocols is analogous to the integration of various
types of moves in Monte Carlo simulation.
It is notable that one of the strongest present
proponents of combinatorial chemistry in the solid state, Henry Weinberg
at SYMYX Technologies, has taught graduate statistical mechanics
for the last twenty-five years!  
Applications of combinatorial chemistry abound
in the fields of catalysis, sensors, coatings, microelectronics, 
biotechnology, and human therapeutics.
Hopefully,
statistical mechanics will have a significant role to play in shaping
these new methods of materials design.

\section*{ACKNOWLEDGMENT}
It is a pleasure to acknowledge the contributions of my collaborators
Leonard D. Bogarad, Marco Falcioni, and Taison Tan.  This work
was supported by the National Science Foundation.

\addcontentsline{toc}{toc}{References\hfill}
\bibliography{advances}

\begin{thebibliography}{75}
\expandafter\ifx\csname natexlab\endcsname\relax\def\natexlab#1{#1}\fi

\bibitem[Agrawal \emph{et~al.}, 1998]{nsf5_15}
Agrawal, A., Eastman, Q.~M., and Schatz, D.~G., Transposition Mediated by RAG1
  and RAG2 and its Implications for the Evolution of the Immune System.
  \emph{Nature} \textbf{394}, 744--751 (1998).

\bibitem[Akporiaye \emph{et~al.}, 1998]{combi_17}
Akporiaye, D.~E., Dahl, I.~M., Karlsson, A., and Wendelbo, R., Combinatorial
  Approach to the Hydrothermal Synthesis of Zeolites. \emph{Angew. Chem. Int.
  Ed.} \textbf{37}, 609--611 (1998).

\bibitem[Bogarad and Deem, 1999]{Deem3}
Bogarad, L.~D., and Deem, M.~W., A Hierarchical Approach to Protein Molecular
  Evolution. \emph{Proc. Natl. Acad. Sci. USA} \textbf{96}, 2591--2595 (1999).

\bibitem[Bratley \emph{et~al.}, 1994]{LDS1}
Bratley, P., Fox, B.~L., and Niederreiter, H., Algorithm-738-Programs To
  Generate {N}iederreiter's {L}ow-{D}iscrepancy {S}equences. \emph{ACM Trans.
  Math. Software} \textbf{20}, 494--495 (1994).

\bibitem[Brice{\~n}o \emph{et~al.}, 1995]{combi_22}
Brice{\~n}o, G., Chang, H., Sun, X., Schultz, P.~G., and Xiang, X.-D., A Class
  of Cobalt Oxide Magnetoresistance Materials Discovered with Combinatorial
  Synthesis. \emph{Science} \textbf{270}, 273--275 (1995).

\bibitem[Burgess \emph{et~al.}, 1996]{combi_31}
Burgess, K., Lim, H.-J., Porte, A.~M., and Sulikowski, G.~A., New Catalysts and
  Conditions for a {C-H} Insertion Reaction Identified by High Throughput
  Catalyst Screening. \emph{Angew. Chem. Int. Ed.} \textbf{35}, 220--222
  (1996).

\bibitem[Cole \emph{et~al.}, 1996]{combi_18}
Cole, B.~M., Shimizu, K.~D., Krueger, C.~A., Harrity, J. P.~A., Snapper, M.~L.,
  and Hoveyda, A.~H., Discovery of Chiral Catalysts through Ligand Diversity:
  {Ti}-Catalyzed Enantioselective Addition of {TMSCN} to \emph{meso} Epoxides.
  \emph{Angew. Chem. Int. Ed.} \textbf{35}, 1668--1671 (1996).

\bibitem[Cong \emph{et~al.}, 1999]{combi_7}
Cong, P., Doolen, R.~D., Fan, Q., Giaquinta, D.~M., Guan, S., McFarland, E.~W.,
  Poojary, D.~M., Self, K., Turber, H.~W., and Weinberg, W.~H., High-Throughput
  Synthesis and Screening of Combinatorial Heterogeous Catalyst Libraries.
  \emph{Angew. Chem. Int. Ed.} \textbf{38}, 484--488 (1999).

\bibitem[Cowell \emph{et~al.}, 1999]{Kepler}
Cowell, L.~G., Kim, H.~J., Humaljoki, T., Berek, C., and Kepler, T.~B.,
  Enhanced Evolvability in Immunoglobulin V Genes under Somatic Hypermutation.
  \emph{J. Mol. Evol.} \textbf{49}, 23--26 (1999).

\bibitem[Crameri \emph{et~al.}, 1998]{len8}
Crameri, A., Raillard, S.~A., Bermudez, E., and Stemmer, W. P.~C., DNA
  Shuffling of a Family of Genes from Diverse Species Accelerates Directed
  Evolution. \emph{Nature} \textbf{391}, 288--291 (1998).

\bibitem[Danielson \emph{et~al.}, 1988]{combi_5}
Danielson, E., Devenney, M., Giaquinta, D.~M., Golden, J.~H., Haushalter,
  R.~C., McFarland, E.~W., Poojary, D.~M., Reaves, C.~M., Weinberg, W.~H., and
  Wu, X.~D., A Rare-Earth Phosphor Containing One-Dimensional Chains Identified
  Through Combinatorial Methods. \emph{Science} \textbf{279}, 837--839 (1988).

\bibitem[Danielson \emph{et~al.}, 1997]{combi_11}
Danielson, E., Golden, J.~H., McFarland, E.~W., Reaves, C.~M., Weinberg, W.~H.,
  and Wu, X.~D., A Combinatorial Approach to the Discovery and Optimization of
  Luminescent Materials. \emph{Nature} \textbf{389}, 944--948 (1997).

\bibitem[Davidson \emph{et~al.}, 1995]{len27}
Davidson, E.~H., Peterson, K.~J., and Cameron, R.~A., Origin of Bilaterian Body
  Plans---Evolution of Developmental Regulatory Mechanisms. \emph{Science}
  \textbf{270}, 1319--1325 (1995).

\bibitem[de~Pablo \emph{et~al.}, 1992]{dePablo}
de~Pablo, J.~J., Laso, M., and Suter, U.~W., Estimation of the Chemical
  Potential of Chain Molecules by Simulation. \emph{J. Chem. Phys.}
  \textbf{96}, 6157 (1992).

\bibitem[Dickinson and Walt, 1997]{combi_24}
Dickinson, T.~A., and Walt, D.~R., Generating Sensor Diversity through
  Combinatorial Polymer Synthesis. \emph{Anal. Chem.} \textbf{69}, 3413--3418
  (1997).

\bibitem[Dowson \emph{et~al.}, 1997]{nsf5_11}
Dowson, C.~G., Barcus, V., King, S., Pickerill, P., Whatmore, A., and Yeo, M.,
  Horizontal Gene Transfer and the Evolution of Resistance and Virulence
  Determinants in {\em Steptococcus}. \emph{J. Appl. Micro. Biol. Symposium
  Supplement} \textbf{83}, 42S--51S (1997).

\bibitem[Falcioni and Deem, 1999]{Deem1}
Falcioni, M., and Deem, M.~W., A Biased Monte Carlo Scheme for Zeolite
  Structure Solution. \emph{J. Chem. Phys.} \textbf{110}, 1754--1766 (1999).

\bibitem[Falcioni and Deem, 2000]{Deem2000}
Falcioni, M., and Deem, M.~W., Library Design in Combinatorial Chemistry by
  Monte Carlo Methods. \emph{Phys. Rev. E} \textbf{61}, 5948--5952 (2000).

\bibitem[Fisch \emph{et~al.}, 1996]{len24}
Fisch, I., Kontermann, R.~E., Finnern, R., Hartley, O., Solergonzalez, A.~S.,
  Griffiths, A.~D., and Winter, G., A Strategy of Exon Shuffling for Making
  Large Peptide Repertoires Displayed on Filamentous Bacteriophage. \emph{Proc.
  Natl. Acad. Sci. USA} \textbf{93}, 7761--7766 (1996).

\bibitem[Frenkel and Smit, 1992]{SmitV}
Frenkel, D., and Smit, B., Unexpected Length Dependence of the Solubility of
  Chain Molecules. \emph{Mol. Phys.} \textbf{75}, 983 (1992).

\bibitem[Frenkel and Smit, 1996]{Frenkel_book}
Frenkel, D., and Smit, B., Understanding Molecular Simulation: From Algorithms
  to Applications. Academic Press, San Diego, 1996.

\bibitem[Frenkel \emph{et~al.}, 1992]{Frenkel}
Frenkel, D., Mooij, C. G. A.~M., and Smit, B., Novel Scheme to Study Structural
  and Thermal Properties of Continuously Deformable Molecules. \emph{J. Phys.:
  Condens. Matter} \textbf{4}, 3053 (1992).

\bibitem[Geyer, 1991]{Geyer}
Geyer, C.~J., Markov Chain {Monte} {Carlo} Maximum Likelihood. in
  \emph{Computing Science and Statistics: Proceedings of the 23rd Symposium on
  the Interface}, American Statistical Association, New {York}, 1991, pp.
  156--163.

\bibitem[Gilbert, 1978]{len28}
Gilbert, W., Why Genes in Pieces? \emph{Nature} \textbf{271}, 501 (1978).

\bibitem[Gilbert \emph{et~al.}, 1997]{len29}
Gilbert, W., DeSouza, S.~J., and Long, M., Origin of Genes. \emph{Proc. Natl.
  Acad. Sci. USA} \textbf{94}, 7698--7703 (1997).

\bibitem[Goffin and Ghuysen, 1998]{Goffin}
Goffin, C., and Ghuysen, J.-M., Multimodular Penicillin-Binding Proteins: An
  Enigmatic Family of Orthologs and Paralogs. \emph{Microbiol. Mol. Biol}
  \textbf{62}, 1079--1093 (1998).

\bibitem[Helmkamp and Davis, 1995]{Davis95}
Helmkamp, M.~M., and Davis, M.~E., Synthesis of Porous Silicates. \emph{Ann.
  Rev. Mater. Sci.} \textbf{25}, 161--192 (1995).

\bibitem[Hobbs \emph{et~al.}, 1997]{nsf5_3}
Hobbs, M.~M., Seiler, A., Achtman, M., and Cannon, J.~G., Microevolution within
  a Clonal Population of Pathogenic Bacteria: Recombination, Gene Duplication
  and Horizontal Genetic Exchange in the {\em opa} gene family of {\em
  Neisseria Meningitidis}. \emph{Molec. Microbiol.} \textbf{12}, 171--180
  (1997).

\bibitem[Jim{\'e}nez-Monta{\~n}o \emph{et~al.}, 1996]{Montano}
Jim{\'e}nez-Monta{\~n}o, M.~A., de~la Mora-Bas{\'a}{\~n}ez, C.~R., and
  P{\"o}schel, T., The Hypercube Structure of the Genetic Code Explains
  Conservative and Non-Conservative Amino Acid Substitutions \emph{in vivo} and
  \emph{in vitro}. \emph{BioSystems} \textbf{39}, 117--125 (1996).

\bibitem[Kamtekar \emph{et~al.}, 1993]{len19}
Kamtekar, S., Schiffer, J.~M., Xiong, H.~Y., Babik, J.~M., and Hecht, M.~H.,
  Protein Design by Binary Patterning of Polar and Nonpolar Amino Acids.
  \emph{Science} \textbf{262}, 1680--1685 (1993).

\bibitem[Kauffman and Levin, 1987]{len14}
Kauffman, S., and Levin, S., Towards a General Theory of Adaptive Walks on
  Rugged Landscapes. \emph{J. Theor. Biol.} \textbf{128}, 11--45 (1987).

\bibitem[Kauffman, 1993]{len15}
Kauffman, S.~A., The Origins of Order. Oxford University Press, New York,
1993.

\bibitem[Kauffman and MacReady, 1995]{len16}
Kauffman, S.~A., and MacReady, W.~G., Search Strategies for Applied Molecular
  Evolution. \emph{J. Theor. Biol.} \textbf{173}, 427--440 (1995).

\bibitem[Kepler, 1997]{nsf5_7}
Kepler, T.~B., Codon Bias and Plasticity in Immunoglobulins. \emph{Mol. Biol.
  Evol.} \textbf{14}, 637--643 (1997).

\bibitem[Kepler and Bartl, 1998]{Kepler2}
Kepler, T.~B., and Bartl, S., Plasticity Under Somatic Mutation in Antigen
  Receptors. \emph{Curr. Top. Micrbiol.} \textbf{229}, 149--162 (1998).

\bibitem[Kepler and Perelson, 1998]{Perelson}
Kepler, T.~B., and Perelson, A.~S., Drug Concentration Heterogeneity
  Facilitates the Evolution of Drug Resistance. \emph{Proc. Natl. Acad. Sci.
  USA} \textbf{95}, 11514--11519 (1998).

\bibitem[Lawrence, 1997]{len12}
Lawrence, J.~G., Selfish Operons and Speciation by Gene Transfer. \emph{Trends
  Microbiol.} \textbf{5}, 355--359 (1997).

\bibitem[Maeshiro and Kimura, 1998]{len31}
Maeshiro, T., and Kimura, M., Role of Robustness and Changeability on the
  Origin and Evolution of Gentic Codes. \emph{Proc. Natl. Acad. Sci. USA}
  \textbf{95}, 5088--5093 (1998).

\bibitem[Maiden, 1998]{Maiden}
Maiden, M. C.~J., Horizontal Genetic Exchange, Evolution, and Spread of
  Antibiotic Resistance in Bacertia. \emph{Clin. Infect. Dis.} \textbf{27},
  S12--S20 (1998).

\bibitem[Mandecki, 1990]{len25}
Mandecki, W., A Method for Construction of Long Randomized Open Reading Frames
  and Polypeptides. \emph{Protein Eng.} \textbf{3}, 221--226 (1990).

\bibitem[Marinari \emph{et~al.}, 1998]{Parisi}
Marinari, E., Parisi, G., and Ruiz-{Lorenzo}, J., Numerical Simulations of Spin
  Glass Systems. in \emph{Spin Glasses and Random Fields} (A.~Young, ed.),
  World Scientific, Singapore, volume~12 of \emph{Directions in Condensed
  Matter Physics}, 1998, pp. 59--98.

\bibitem[Massova and Mobashery, 1999]{Massova}
Massova, I., and Mobashery, S., Structural and Mechanistic Aspects of Evolution
  of $\beta$-Lactamases and Penicillin-Binding Proteins. \emph{Curr. Pharm.
  Design} \textbf{5}, 929--937 (1999).

\bibitem[Medeiros, 1997]{Medeiros}
Medeiros, A.~A., Evolution and Dissemination of $\beta$-Lactamases Accelerated
  by Generation of $\beta$-Lactam Antiobiotics. \emph{Clin. Infect. Dis.}
  \textbf{24}, S19--S45 (1997).

\bibitem[Menger \emph{et~al.}, 1995]{combi_16}
Menger, F.~M., Eliseev, A.~V., and Migulin, V.~A., Phosphatase Catalysis
  Develped \emph{via} Combinatorial Organic Chemistry. \emph{J. Org. Chem.}
  \textbf{60}, 6666--6667 (1995).

\bibitem[Miller and Orgel, 1974]{len21}
Miller, S., and Orgel, L., The Origin of Life on Earth. Prentice Hall, London,
1974.

\bibitem[Moore \emph{et~al.}, 1997]{len10}
Moore, J.~C., Jin, H.-M., Kuchner, O., and Arnold, F.~H., Strategies for the
  \emph{in-vitro} Evolution of Protein Function---Enzyme Evolution by Random
  Recombination of Improved Sequences. \emph{J. Mol. Bio.} \textbf{272},
  336--347 (1997).

\bibitem[Moran \emph{et~al.}, 1999]{Moran}
Moran, J.~V., DeBerardinis, R.~J., and Kazazian, H.~H., Exon Shuffling by L1
  Retrotransposition. \emph{Science} \textbf{283}, 1530--1534 (1999).

\bibitem[Netzer and Hartl, 1997]{len30}
Netzer, W.~J., and Hartl, F.~U., Recombination of Protein Domains Facilitated
  by Co-Translational Folding in Eukaryotes. \emph{Nature} \textbf{388},
  343--349 (1997).

\bibitem[Niederreiter, 1992]{LDS2}
Niederreiter, H., Random Number Generation and Quasi-{M}onte {C}arlo Methods.
  Society for Industrial and Applied Mathematics, Philadelphia, 1992.

\bibitem[Novet \emph{et~al.}, 1995]{Novet95}
Novet, T., Johnson, D.~C., and Fister, L., Interfaces, Interfacial Reactions
  and Superlattice Reactants. \emph{Adv. Chem. Ser.} \textbf{245}, 425--469
  (1995).

\bibitem[Ostermeier \emph{et~al.}, 1999{\natexlab{a}}]{Benkovic1}
Ostermeier, M., Nixon, A.~E., and Benkovic, S.~J., Incremental Truncation as a
  Strategy in the Engineering of Novel Biocatalysts. \emph{Bioorganic \& Med.
  Chem.} \textbf{7}, 2139--2144 (1999{\natexlab{a}}).

\bibitem[Ostermeier \emph{et~al.}, 1999{\natexlab{b}}]{Benkovic2}
Ostermeier, M., Shim, J.~H., and Benkovic, S.~J., A Combinatorial Approach to
  Hybrid Enzymes Independent of DNA Homology. \emph{Nature Biotech.}
  \textbf{17}, 1205--1209 (1999{\natexlab{b}}).

\bibitem[Patten \emph{et~al.}, 1997]{len11}
Patten, P.~A., Howard, R.~J., and Stemmer, W. P.~C., Applications of DNA
  Shuffling to Pharmaceuticals and Vaccines. \emph{Curr. Opin. Biotech.}
  \textbf{8}, 724--733 (1997).

\bibitem[Pennisi, 1998]{len13}
Pennisi, E., How the Genome Readies Itself for Evolution. \emph{Science}
  \textbf{281}, 1131--1134 (1998).

\bibitem[Perelson and Macken, 1995]{len17}
Perelson, A.~S., and Macken, C.~A., Protein Evolution on Partially Correlated
  Landscapes. \emph{Proc. Natl. Acad. Sci. USA} \textbf{92}, 9657--9661 (1995).

\bibitem[Pirrung, 1997]{combi_2a}
Pirrung, M.~C., Spatially Addressable Combinatorial Libraries. \emph{Chem.
  Rev.} \textbf{97}, 473--488 (1997).

\bibitem[Plasterk, 1998]{nsf5_13}
Plasterk, R., V(D)J Recombination: Ragtime Jumping. \emph{Nature} \textbf{394},
  718--719 (1998).

\bibitem[Reddington \emph{et~al.}, 1998]{combi_21}
Reddington, E., Sapienza, A., Gurau, B., Viswanathan, R., Sarangapani, S.,
  Smotkin, E.~S., and Mallouk, T.~E., Combinatorial Electrochemistry: {A}
  Highly Parallel, Optical Screening Method for Discovery of Better
  Electrocatalysts. \emph{Science} \textbf{280}, 1735--1737 (1998).

\bibitem[Riddle \emph{et~al.}, 1997]{len20}
Riddle, D.~S., Santiago, J.~V., Brayhall, S.~T., Doshi, N., Grantcharova,
  V.~P., Yi, Q., and Baker, D., Functional Rapidly Folding Proteins from
  Simplified Amino Acide Sequences. \emph{Nature Struct. Biol.} \textbf{4},
  805--809 (1997).

\bibitem[Schuster and Stadler, 1998]{len22}
Schuster, P., and Stadler, P.~F., Sequence Redundancy in Biopolymers: A Study
  on RNA and Protein Structures. in \emph{Viral Regulatory Structures and their
  Degeneracy} (G.~Myers, ed.), Addison-Wesley, New York, 1998, pp. 163--186.

\bibitem[Sedgewick, 1988]{Sedgewick}
Sedgewick, R., Algorithms. Addison-Wesley, New York, 2nd ed., 1988.

\bibitem[Shapiro, 1992]{Shapiro}
Shapiro, J.~A., Natural Genetic Engineering in Evolution. \emph{Genetica}
  \textbf{86}, 99--111 (1992).

\bibitem[Shapiro, 1997]{Shapiro2}
Shapiro, J.~A., Genome Organization, Natural Genetic Engineering and Adaptive
  Mutation. \emph{Trends in Genetics} \textbf{13}, 98--104 (1997).

\bibitem[Smit and Maesen, 1995]{review_83}
Smit, B., and Maesen, T. L.~M., Commensurate `Freezing' of Alkanes in the
  Channels of a Zeolite. \emph{Nature} \textbf{374}, 42 (1995).

\bibitem[Smith \emph{et~al.}, 1996]{Smith96}
Smith, D.~S., Creadon, G., Jena, P.~K., Portanova, J.~P., Kotzin, B.~L., and
  Wysocki, L.~J., Di- and Trinucleotide Target Preferences of Somatic
  Mutagenesis in Normal and Autoreactive B Cells. \emph{J. Immunol.}
  \textbf{156}, 2642--2652 (1996).

\bibitem[Stemmer, 1994]{len7}
Stemmer, W. P.~C., Rapid Evolution of a Protein \emph{in-vitro} by DNA
  Shuffling. \emph{Nature} \textbf{370}, 389--391 (1994).

\bibitem[Stemmer \emph{et~al.}, 1995]{len26}
Stemmer, W. P.~C., Crameri, A., Ha, K.~D., Brennan, T.~M., and Heyneker, H.~L.,
  Single-Step Assembly of a Gene and Entire Plasmid from Large Numbers of
  Oligodeoxyribonucleotides. \emph{Gene} \textbf{164}, 49--53 (1995).

\bibitem[Syvanen, 1997]{nsf5_20}
Syvanen, M., Horizontal Gene Transfer: Evidence and Possible Consequences.
  \emph{Annu. Rev. Genet} \textbf{28}, 237--261 (1997).

\bibitem[van Dover \emph{et~al.}, 1998]{combi_8}
van Dover, R.~B., Schneemeyer, L.~F., and Fleming, R.~M., Discovery of a Useful
  Thin-Film Dielectric Using a Composition-Spread Approach. \emph{Nature}
  \textbf{392}, 162--164 (1998).

\bibitem[Volkenstein, 1994]{Volkenstein}
Volkenstein, M.~V., Physical Approaches to Biological Evolution.
  Springer-Verlag, New York, 1994.

\bibitem[Wang \emph{et~al.}, 1998]{combi_10}
Wang, J., Yoo, Y., Gao, C., Takeuchi, I., Sun, X., Chang, H., Xiang, X.-D., and
  Schultz, P.~G., Identification of a Blue Photoluminescent Composite Material
  from a Combinatorial Library. \emph{Science} \textbf{279}, 1712--1714 (1998).

\bibitem[Weinberg \emph{et~al.}, 1998]{combi_15}
Weinberg, W.~H., Jandeleit, B., Self, K., and Turner, H., Combinatorial Methods
  in Homogeneous and Heterogeneous Catalysis. \emph{Curr. Opin. Chem. Bio.}
  \textbf{3}, 104--110 (1998).

\bibitem[Xiang \emph{et~al.}, 1995]{combi_4}
Xiang, X.-D., Sun, X., Brice{\~n}o, G., Lou, Y., Wang, K.-A., Chang, H.,
  Wallace-Freedman, W.~G., Chang, S.-W., and Schultz, P.~G., A Combinatorial
  Approach to Materials Discovery. \emph{Science} \textbf{268}, 1738--1740
  (1995).

\bibitem[Zhang \emph{et~al.}, 1997]{len9}
Zhang, J.-H., Dawes, G., and Stemmer, W. P.~C., Directed Evolution of a
  Fucosidase from a Galactosidase by DNA Shuffling and Screening. \emph{Proc.
  Natl. Acad. Sci. USA} \textbf{94}, 4504--4509 (1997).

\bibitem[Zones \emph{et~al.}, 1998]{Zones98}
Zones, S.~I., Nakagawa, Y., Lee, G.~S., Chen, C.~Y., and Yuen, L.~T., Searching
  for New High Silica Zeolites through a Synergy of Organic Templates and Novel
  Inorganic Conditions. \emph{Micropor. Mesopor. Mat.} \textbf{21}, 199--211
  (1998).

\end{thebibliography}

\end{document}